\pdfoutput=1
\documentclass{aa}

\usepackage{graphicx}
\usepackage{txfonts}
\usepackage{amsmath}
\newcommand{\Msun}{\mbox{M$_{\odot}$}}
\newcommand{\Rsun}{\mbox{R$_{\odot}$}}
\newcommand{\Lsun}{\mbox{L$_{\odot}$}}
\newcommand{\kms}{\mbox{km\,s$^{-1}$}}
\newcommand{\ms}{\mbox{m\,s$^{-1}$}}

\begin{document}

\title{Time-series Doppler imaging of the red giant
\object{HD~208472}\thanks{Based on data obtained with
the STELLA robotic telescope in Tenerife, an AIP
facility jointly operated by AIP and IAC, and the
Potsdam Automatic Photoelectric Telescopes (APT) in
Arizona, jointly operated by AIP and Fairborn
Observatory.}}

\subtitle{Active longitudes and differential rotation}

\author{O. \"Ozdarcan\inst{1,2}, T. A. Carroll\inst{1},
A. K\"unstler\inst{1}, K. G. Strassmeier\inst{1},
S. Evren\inst{2}, M. Weber\inst{1} \and
T. Granzer\inst{1}}

\authorrunning{O.~\"Ozdarcan et al.}

\institute{Leibniz Institute for Astrophysics Potsdam
(AIP), An der Sternwarte 16, 14482 Potsdam, Germany\\
\email{orkun.ozdarcan@[ege.edu.tr;gmail.com];
[tcarroll;kstrassmeier]@aip.de}
\and
Ege University, Faculty of Science, Astronomy and Space
Sciences Department, 35100 Bornova - \.Izmir, Turkey}

\date{Received ---------; accepted ---------}

\abstract
{HD~208472 is among the most active RS~CVn binaries with 
cool starspots. Decade-long photometry has shown that the 
spots seem to change their longitudinal appearance with a 
period of about six years, coherent with brightness 
variations.}
{Our aim is to spatially resolve the stellar surface of 
\object{HD~208472} and relate the photometric results to 
the true longitudinal and latitudinal spot appearance. 
Furthermore, we investigate the surface differential rotation
pattern of the star.}
{We employed three years of high-resolution
spectroscopic data with a high signal-to-noise ratio (S/N)\ from the STELLA robotic observatory and 
determined new and more precise stellar physical parameters.
Precalculated synthetic spectra were fit to each of these 
spectra, and we provide new spot-corrected orbital elements. 
A sample of 34 absorption lines per spectrum was used to 
calculate mean line profiles with a S/N of several hundred.
A total of 13 temperature Doppler images were reconstructed 
from these line profiles with the inversion code $iMap$. 
Differential rotation was investigated by cross-correlating 
successive Doppler images in each observing season.}
{Spots on HD~208472 are distributed preferably at high 
latitudes and less frequently around mid-to-low 
latitudes. No polar-cap like structure is seen at any 
epoch. We observed a flip-flop event between 2009 
and 2010, manifested as a flip of the spot activity from 
phase 0.0 to phase 0.5, while the overall 
brightness of the star continued to increase and reached an
all-time  maximum in 2014. Cross-correlation of successive 
Doppler images suggests a solar-like differential rotation
that is $\approx$ 15 times weaker than that of
the Sun.}
% conclusions heading (optional), leave it empty if
%necessary
%{}
{}


\maketitle

%
%________________________________________________________________

\section{Introduction}\label{S1}

Well-sampled photometric monitoring of spotted cool stars 
allows tracing photospheric brightness variations that are caused by 
rotational modulation. Long-term photometric data moreover 
allow investigating spot activity cycles and relate 
inherent changes of the surface spot distribution to 
photometric period variations, both of which are an 
indicator of the underlying surface differential rotation.
Phenomena such as active longitudes \citep{henry95,jetsu96}
and flip-flops of active longitudes \citep{jetsu91,berd_tuo98} 
were revealed by such long-term photometry, but they are far 
from being fully understood. \citet{holz03a,holz03b} 
demonstrated on theoretical grounds that the tidal effects 
of a companion star in a binary system may alter the 
surface distribution of spots by breaking the axial 
symmetry, which could lead to preferred longitudes.
Magnetic dynamo simulations \citep{fluri04,moss05,elstner05}
of the flip-flop phenomenon basically concluded that the 
axisymmetric and the non-axisymmetric (magnetic) field 
components must coexist to produce a flip-flop in 
the first place. These studies highlighted that meridional 
circulation and differential rotation are a prerequisite for 
this type of dynamo but do not yet allow predicting 
the relation between differential rotation and 
flip-flop period, for example. Recently, \citet{rot15a, rot15b} 
showed that there might be ambiguity in interpreting light 
curves of RS CVn binaries in terms of persistent active 
longitudes and ellipsoidal distortion. In this case, some 
part of the light curve might be caused by binarity
effects and not by cool spots.

For the characterization of differential surface rotation, 
time-series Doppler imaging became the most powerful method 
(see, e.g., \citealt{donati97,strass09,kunstler15,kov16}). 
The differential rotation signature is either based on the 
longitudinal cross-correlation of images from successive 
rotation cycles or on the addition of the image shear as an
additional free parameter in the line-profile inversion 
itself. Neither of these variants is free of problems, see 
\citet{dun08} for a comparison, but both were successfully 
applied to a few single and binary stars in different evolutionary 
stages (e.g., \citealt{don03,dun08,kov12}). 
For main-sequence stars, only solar-type differential rotation 
(equator rotates faster than poles) was found so far, while 
anti-solar differential rotation (poles rotate faster than equator)
was found for a few fast-rotating giants \citep{weber05}. 
\citet{kit_rud04} have shown that fast meridional 
circulation in a very deep convection zone with magnetically 
induced thermal inhomogeneities would lead to anti-solar 
differential rotation. However, at least in some cases, different 
authors found contradictory results for one and the same target, 
for example, for the K sub-giant of the RS~CVn binary \object{HR~1099}, 
for which \citet{vogt99} analyzed high-resolution unpolarized 
spectroscopic data and found weak anti-solar differential 
rotation, while \citet{petit04} used high-resolution polarized 
spectra and found solar-type differential rotation. It is mostly
assumed that the spots follow the rotation of the surface, while 
 differential rotation is investigated by photometric spot modeling 
and/or time-series Doppler imaging. However, \citet{korh11} suggested 
that large spot structures could reflect geometric properties of the 
large-scale dynamo and not the surface rotation. Still, the spots
are the tracers we rely on.

\object{HD~208472} (\object{V2075 Cyg}, \object{HIP 108198}, 
$V$=7\fm4, $P_{\rm phot}$=22.4\,d) is a conspicuous target that 
seems to show active spot longitudes and an activity cycle 
with a period of about six years \citep{ozd10} (hereafter paper~I). 
In 1991, W.~Bidelman discovered its strong Ca\,{\sc ii} H and K emission, 
and a number of follow-up studies revealed and confirmed that the system 
is a single-lined spectroscopic binary with a G8III primary component 
on a nearly circular orbit that exhibits rotationally modulated light variations 
with variable light-curve amplitude \citep{henry95,fekel99,strass99,koen_eye02}. 
It was also suggested that the system may be a good candidate for 
Doppler imaging \citep{henry95}. First such attempts for Doppler 
imaging were made by \citet{weber01}, \citet{weber04} and \citet{weber05}, whose initial 
study was based on 70 consecutive nights of spectroscopic observations 
with the NSO McMath-Pierce telescope between 1996 and 1997. They found 
that spots at that time were concentrated at low and intermediate 
latitudes without any structures close to the one visible pole of the star. 
They applied the sheared-image method and found a weak anti-solar 
differential rotation of $\alpha=-0.04\pm0.02,$ which, however, was not 
considered a conclusive result. \citet{erdem09} estimated an $\alpha$ 
parameter of the same amount, but relative to the orbital period, through
photometric spot modeling based on $BVR_{c}I_{c}$ photometry in 2006. 
More comprehensive photometry was presented in our paper~I, where we 
analyzed 17 years of data from Automatic Photoelectric Telescopes (APT). 
A spot-modeling and photometric-period 
analysis led us to a differential rotation coefficient of 
$\alpha=0.004\pm0.010$, which practically indicated a non-detection 
of differential rotation. Furthermore, we found that the longitudinal 
position of spots varied coherently with the mean brightness and seemed 
to exhibit a 6.28\,yr period. This cyclic variation was interpreted 
as a stellar analog of the solar 11-year sunspot cycle.

In the current study, we present new high-resolution time-series 
spectroscopy of the system from 2009--2011 together with contemporaneous 
$V$-band photometry. We redetermine precise atmospheric parameters, 
obtain an improved spectroscopic orbit ,and investigate the evolutionary 
status of the primary star. The unique time-series data from STELLA 
enables us to obtain Doppler images of the star for many successive 
rotational cycles, and for three consecutive years. We first summarize 
the instrumental setup and the data collection and reduction procedures 
in Sect.~\ref{S2}. In Sect.~\ref{S3} we derive stellar atmospheric parameters, 
spectroscopic orbital elements and physical properties of the system. Section~\ref{S4} comprises the Doppler imaging of the 
primary star, including data preparation and a brief description of the 
$iMap$ inversion code. We present a total of 13 surface maps and their 
analysis in Sect.~\ref{S5}. In the final section, we summarize and 
discuss our results.

\section{Observations and data reductions}\label{S2}

Spectroscopic observations were carried out with the robotic STELLA 
facility on Tenerife, Spain, from March 22, 2009 to July 20, 2011. 
All spectra were obtained with the fiber-fed STELLA Echelle Spectrograph 
(SES) with a resolving power of $R$=55,000 corresponding to a two-pixel 
resolution of 0.12\,\AA\ at 6400\,\AA . The spectra cover the wavelength 
range 3900--8800\,\AA . In most cases, a single exposure was taken per 
available night. On some occasions, two successive spectra were obtained. 
Exposure time was set to 7,200\,s during the 2009 observing season, and 
4,500\,s thereafter. The average signal-to-noise ratio (S/N) varied 
between 240 and 160 for the three observing seasons. By 2010, the 
efficiency for coupling the starlight into the fiber was significantly 
improved, and equally good S/N spectra were obtained with a shorter exposure 
time. In total, 229 usable spectra were acquired during three years.

Raw spectra were reduced with the SES data reduction pipeline, which is 
mainly based on IRAF\footnote{The Image Reduction and Analysis Facility 
is hosted by the National Optical Astronomy Observatories in Tucson, Arizona 
at URL iraf.noao.edu.} command-language scripts \citep{weber08}. CCD images 
are corrected for bad pixel and cosmic-ray impacts. Average overscan is 
calculated and subtracted from each image. An already overscan-subtracted 
master bias frame is applied to the images to remove bias levels. Nightly 
obtained flat-field images (around dusk, dawn, and midnight) are used to produce 
a master flat-field image of the related night, and this master image is 
normalized to unity. Then target spectra are divided by this master flat. 
Scattered-light correction is done from the inter-order space, and one-dimensional 
spectra are extracted with the standard IRAF optimal-extraction routine. 
After this, a wavelength calibration is applied by using consecutively recorded 
Th-Ar spectra. Finally, continuum normalization of the extracted spectral 
orders is done by dividing with a flux-normalized synthetic spectrum of 
spectral classification just like HD~208472.

Photometric observations were conducted with the T7 0.75\,m APT $Amadeus$ at Fairborn Observatory in 
southern Arizona \citep{strassmeier97}. Its data were extensively used 
and described in paper~I, and we refer to this paper for 
instrumental and observational details.

\section{Astrophysical parameters}\label{S3}

\subsection{Unspotted brightness}\label{S3.1}

\begin{figure}[!htb]
\includegraphics[angle=0,width=87mm,clip]{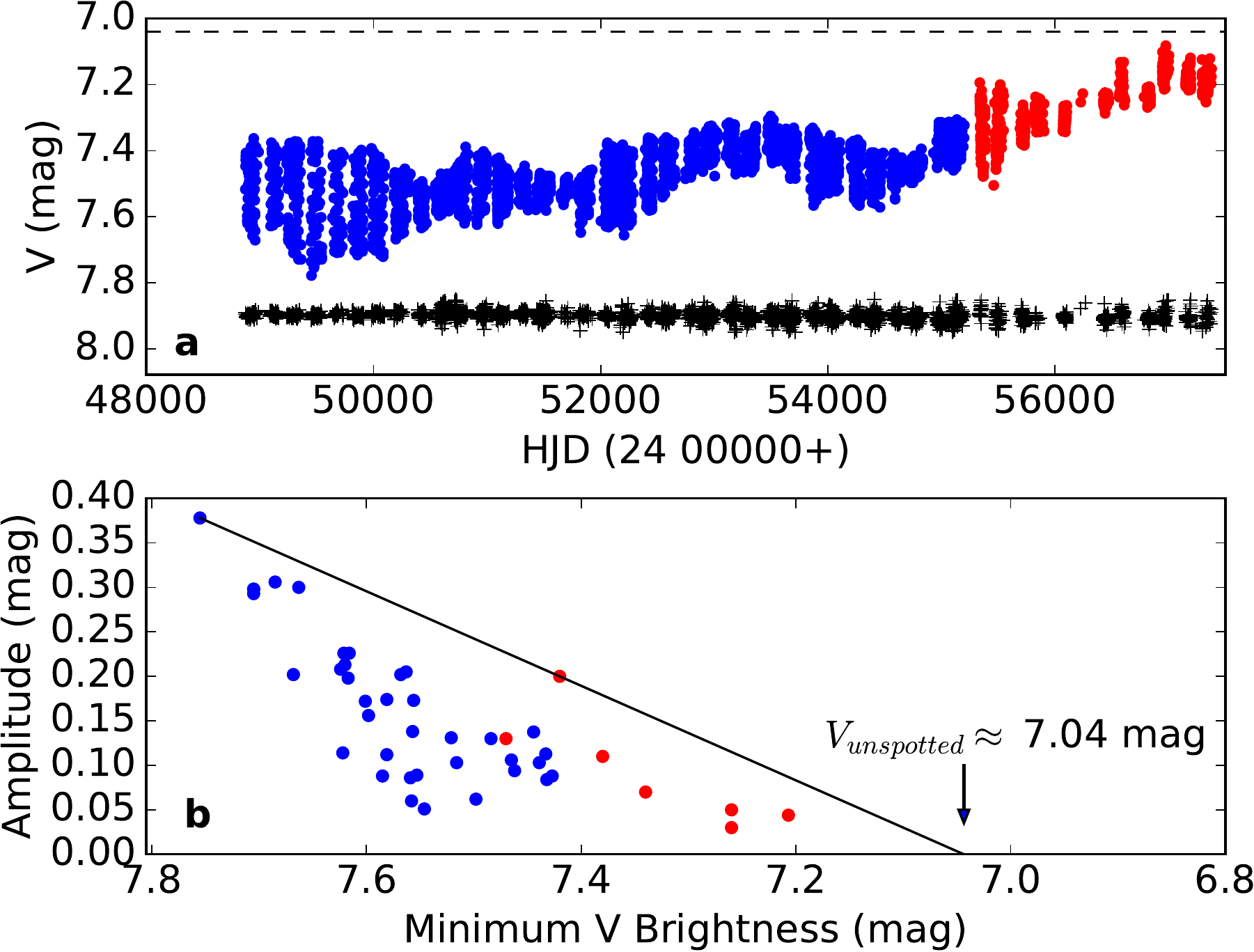}
\caption{\textbf{a.} Long-term $V$-band photometry of
HD~208472. The dashed line denotes the updated unspotted
brightness level. The blue filled dots show data from
paper~I, while the red dots denote the recent data. 
Plus signs show differential magnitudes in the sense 
of check-minus-comparison star. These measurements are shifted
by a constant amount and plotted on a common scale with 
measurements of the variable star.
\textbf{b.} Minimum brightness vs. $V$ amplitude diagram.
The straight line is a linear representation of the upper
envelope of the scatter.}
\label{F1}
\end{figure}

Figure~\ref{F1}a shows all available Johnson $V$-band data of \object{HD\,208472}. 
With the addition of our new APT data for the time range from 2010 to 2015, 
the photometric coverage now extends continuously over 23 years. We determined 
the peak-to-peak light curve amplitudes and minimum brightness levels for each
observing season and found that the system was never brighter than by the end of 
2014, reaching a $V$ magnitude of 7\fm13$\pm$0.01. This is even brighter than 
the predicted unspotted (apparent) brightness of 7\fm323 previously estimated 
in paper~I. Using the minimum brightness vs. amplitude method of \citet{olah97}, 
we find a new value for the unspotted $V$ brightness of 7\fm04 (Fig.~\ref{F1}b).
Using the interstellar reddening obtained in paper~I ($E(B-V)$=0\fm127), we revise 
the intrinsic unspotted $V$ brightness of HD~208472 to 6\fm65$\pm$0.01.

This updated value together with the \emph{Hipparcos} distance of 153\,pc 
\citep{vanlee07} modifies the absolute $V$ magnitude of HD\,208472 to +0\fm73$\pm$0.18. 
Its luminosity is therefore 64$\pm$5 L$_\sun$ (adopting $M_{\rm bol,\odot}$=+4\fm74).

%################ Figure 1 - Photometry

\subsection{Orbital elements}\label{S3.2}

Radial velocities (RV) from STELLA-SES spectra were derived from an 
order-by-order cross correlation with a synthetic template spectrum 
and then averaged. A total of 60 orders out of the 80 available were 
used, and a synthetic template spectrum of a G8 giant was adopted. 
The standard (external) error of a single  STELLA-SES observation of 
\object{HD\,208472} is $\approx$30\,\ms. No systemic zero-point shift 
was added to the data in this paper, but see \citet{strass12} for its 
determination with respect to the CORAVEL reference system. 
The complete data set of 229 RVs is available in electronic form at the
CDS Strasbourg.

\begin{figure*}[!htb]
\centering
%\subfloat[Best representation of uncorrected velocities.]
{\includegraphics[angle=0,scale=0.50,clip=true]{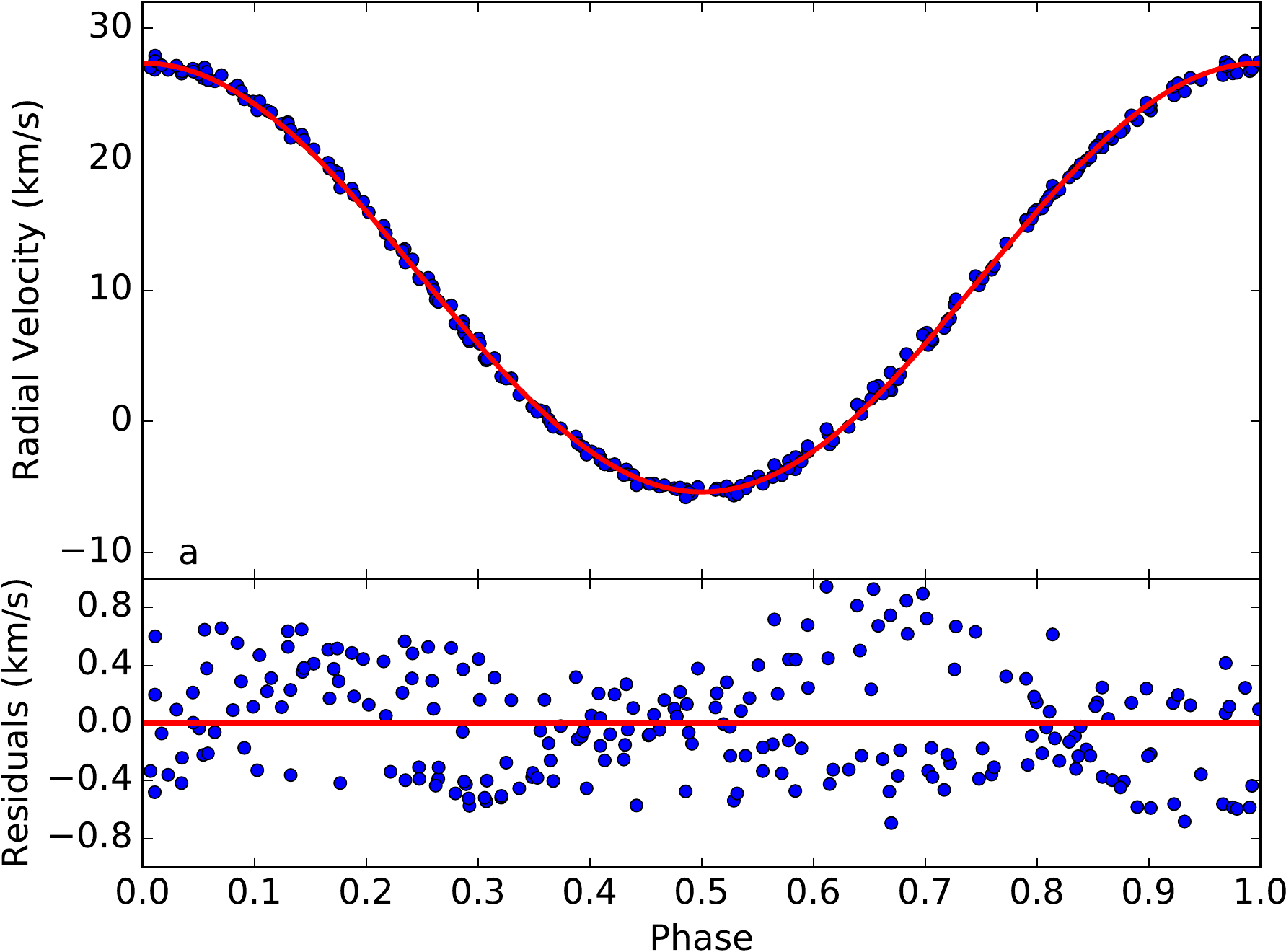}}
%\hfill
%\subfloat[Best representation of spot corrected velocities.]
{\includegraphics[angle=0,scale=0.50,clip=true]{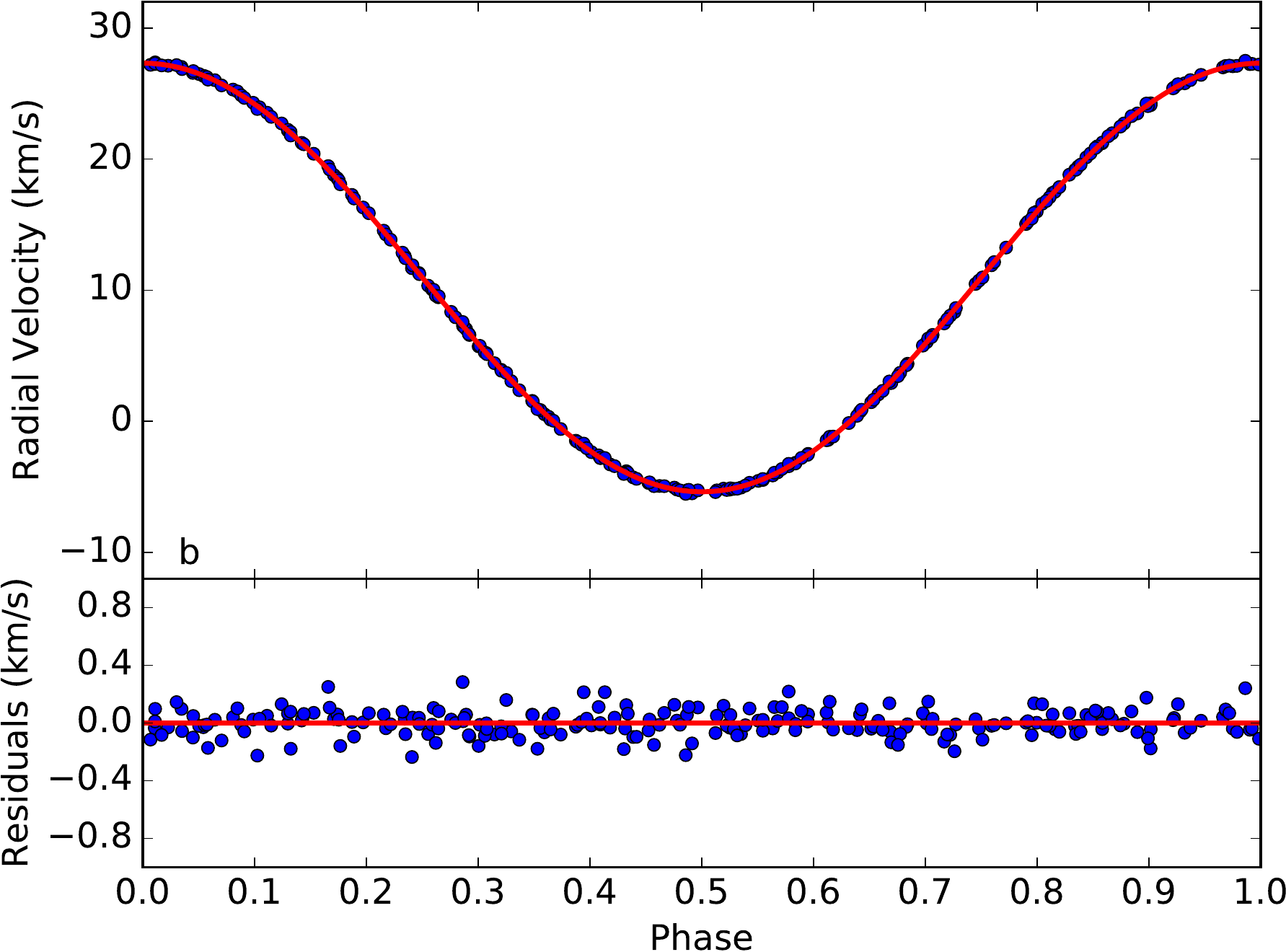}}
\caption{\textbf{a.} Radial velocities (dots) before spot correction and the 
representation of the orbital fit (line). The bottom panel shows the residuals.
\textbf{b.}
Spot-corrected radial velocities and the final fit for the orbital elements. 
The rms of the fit decreased by a factor four with respect to the uncorrected 
velocities.}\label{F2}
\end{figure*}

For an initial orbital solution we adopted the elements from 
\citet{fekel99} as starting values. The preliminary solution showed 
clear and systematic residuals with a full amplitude of up to 1\,\kms\ 
(Fig.~\ref{F2}a). Two anti-phase waves can be identified (lower panel of the figure), 
indicating that two spots or spot groups traversed the stellar disk during 
the time of observation. We applied a discrete Fourier transform to the 
residuals using the Period04 \citep{lenz_breg04} software package and 
searched for significant periods. This was done separately for the 
individual observing season. In 2010, we subdivided the time coverage 
into two data groups because the variation pattern had changed  between 
the first and second half of the season. The resulting periods 
with significance of 5$\sigma$ or more in amplitude are tabulated in 
Table~\ref{T1} together with their amplitudes. The dominant period is 
close to the photometric (= rotational) period in 2010 and 2011, while the other 
periods are just the first, second, and third harmonics of it. In 2009,
half of the photometric period is the most dominant one and the other periods are
the photometric period itself and its other harmonics. This 
confirms that the RV scatter is due to starspots migrating in and out 
of sight and not due to an as yet undiscovered planet.

In the next step, we pre-whitened the RV measurements of every individual 
observing sequence with a least-squares fit from the periods in Table\ref{T1}. 
These data are called the spot-corrected velocities. We then repeated the 
orbital solution and obtained the final elements given in Table~\ref{T2}. 
Figure~\ref{F2}b shows the best orbital fit with respect to the spot-corrected 
RVs and its residuals. We note that the orbital elements remain almost identical, 
but the errors were reduced on average by a factor of 4.

%################ Figure 2 - Radial Velocities

%################ Table 1 - RV residuals frequencies
\begin{table}[!htb]
\begin{flushleft}
\caption{Periodogram results of radial velocity residuals.}\label{T1}
\begin{tabular}{crc}
\hline \hline \noalign{\smallskip}
Year & Period (d) & Amplitude (\kms) \\
\noalign{\smallskip} \hline \noalign{\smallskip}
2009   &  11.127$\pm$0.006    &    0.48$\pm$0.01 \\
       &  22.882$\pm$0.052    &    0.24$\pm$0.01 \\
       &   7.382$\pm$0.018    &    0.07$\pm$0.01 \\
       &  25.322$\pm$0.189    &    0.07$\pm$0.01 \\
\hline \noalign{\smallskip}
2010-A &  22.642$\pm$0.111    &    0.39$\pm$0.02 \\
       &  11.368$\pm$0.034    &    0.32$\pm$0.02 \\
       &   7.375$\pm$0.020    &    0.23$\pm$0.02 \\
       &   5.580$\pm$0.030    &    0.09$\pm$0.02 \\
\hline \noalign{\smallskip}
2010-B &  22.215$\pm$0.170    &    0.33$\pm$0.02 \\
       &   7.484$\pm$0.034    &    0.19$\pm$0.02 \\
       &  10.857$\pm$0.081    &    0.17$\pm$0.02 \\
\hline \noalign{\smallskip}
2011   &  23.059$\pm$0.375    &    0.28$\pm$0.03 \\
\hline \noalign{\smallskip}
\end{tabular}
\end{flushleft}
\end{table}

%################ Table 2 - Orbital Elements
\begin{table}[!tbh]
\begin{flushleft}
\caption{Spectroscopic orbital elements.}\label{T2}
\scalebox{0.95}{
\begin{tabular}{lcc}
\hline \hline \noalign{\smallskip}
                         & Uncorrected             & Spot corrected\\
\noalign{\smallskip} \hline \noalign{\smallskip}
$P_{\rm orb}$ (days)     &   22.61942$\pm$0.00066  &   22.61940$\pm$0.00015 \\
$T_{\rm 0}$ (HJD2449+)   &    252.414$\pm$0.18     &    252.419$\pm$0.041   \\
$\gamma$ (\kms)          &     10.968$\pm$0.012    &     10.974$\pm$0.003   \\
$K_{1}$ (\kms)           &     16.355$\pm$0.035    &     16.345$\pm$0.008   \\
$e$ (adopted)            &           0             &           0            \\
$\omega$                 &          \dots           &         \dots           \\
$a_1\sin i$ (10$^6$ km)  &      5.087$\pm$0.011    &      5.084$\pm$0.003   \\
$f(m)$ (M$_{\sun}$)      &    0.01025$\pm$0.00006  &    0.01023$\pm$0.00001 \\
fit rms (\ms )           &          380            &           88           \\
\noalign{\smallskip} \hline
\end{tabular}
}

Note. The elements are for the unaltered RV data (column ``Uncorrected'') 
and for starspot-jitter removed data (column ``Spot corrected'').
\end{flushleft}
\end{table}

\subsection{Astrophysical parameters}\label{S3.3}

Effective temperature, surface gravity, metallicity and microturbulence 
were determined from each individual SES spectrum with the PARSES pipeline 
\citep{cal04,cal06,jov13}. This fits precalculated grids of synthetic 
spectra to selected wavelength regions of the observed spectra. Grids of 
synthetic spectra were calculated from MARCS model atmospheres in local thermodynamic
equilibrium (LTE) \citep{gustaffson08}. The pipeline searches for the best-fit synthetic 
spectrum by minimizing the difference between the observed and computed spectrum and 
then adopts its parameters as its best solution. In addition to the above four 
atmospheric parameters, a line-broadening parameter consisting of the combined 
effect of $v\sin i$ and macroturbulence was adopted as a free parameter during 
the fitting process. According to the atmospheric properties of the star, an 
assumed macroturbulence of 5\,\kms\ was adopted from \citet{graybook} and kept 
fixed.

\begin{figure}[!htb]
\includegraphics[angle=0,scale=0.47,clip=true]{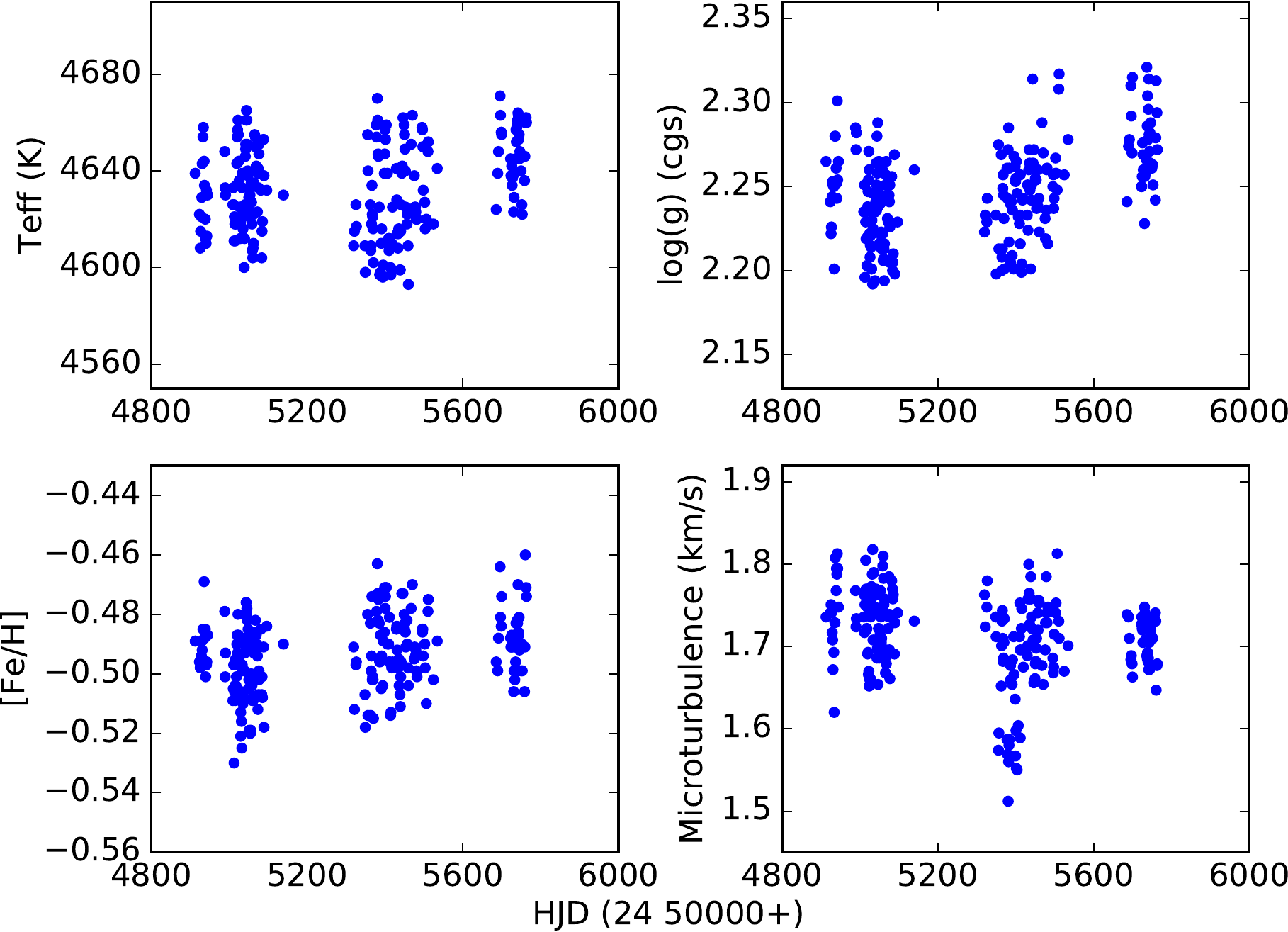}
\caption{PARSES spectrum-synthesis results of four atmospheric parameters for 
each individual SES  observation.}\label{F3}
\end{figure}

The PARSES results for 218 SES spectra are plotted as a function of Julian date 
in Fig.~\ref{F3}. An effective temperature of 4,630\,K, $\log g$=2.25, [Fe/H]=--0.49, 
and a microturbulence of 1.7\,\kms\ compromise the best-fit averages. We excluded 
11 of the 229 spectra that grossly deviated from the bulk, possibly because of
problems arising from the continuum level. From the remaining individual values, 
we took unweighted averages and adopted these as our final atmospheric parameters 
in Table~\ref{T3}. Their rms deviations are estimated to be the respective internal errors.

%################ Figure 3 - PARSES results

%18.9$\pm$0.5~\kms\ (rms)

The PARSES pipeline returns average $v\sin i$ values between 18 and 24\,\kms, 
depending on the assumed radial-tangential macro- and microturbulence and the 
wavelength range selected. A much more accurate $v\sin i$ value of 19.4$\pm$0.2~\kms\ 
is possible from our Doppler imaging analysis (see Sect.~\ref{S4}), in good agreement 
with 19.7~\kms\ given by \citet{fekel99}. The radius of the star derived with the 
Stefan-Boltzmann law is then 12.5$\pm$0.5\,\Rsun , a typical value for an early-K 
giant (solar $T_{\rm eff}$ and $M_{\rm bol}$ were adopted as 5780\,K and 4\fm74, 
respectively). The bolometric correction is adopted from \citet{flower96}. Combination 
of this radius and the $v\sin i$ leads to an inclination angle of 44$^{\circ}$. 
Simple propagation of the internal errors of the measured and calculated parameters 
results in an unrealistically small uncertainty for the inclination (just a few degrees). 
In particular for active giant stars, uncertainties on calculated radius and 
luminosity are expected to be much larger, hence we estimate $\pm 10^{\circ}$ for 
the range of the most likely inclination.

%################ Table 3 - Astrophysical parameters
\begin{table}[!htb]
\caption{Astrophysical parameters of \object{HD~208472}.
Errors quoted are internal errors.}
\label{T3}
%Bolometric correction and macro-turbulence turbulence
%values were adopted from \citet{flower96} and
%\citet{graybook}, respectively.}\label{T3}
\begin{center}
 \begin{tabular}{ll}
\hline \hline \noalign{\smallskip}
  Parameter                & Value                                    \\
\noalign{\smallskip} \hline \noalign{\smallskip}
  Spectral Type            & K1 III                                  \\
  $V_{0}$                  & 6\fm65                                \\
  Bolometric correction    & --0\fm51                              \\ % from Flower (1996)
  $M_{V}$                  & 0\fm73$\pm$0.18                      \\
  $M_{\rm bol}$            & 0\fm22                               \\
  $v\sin i$                & 19.4$\pm$0.2~\kms                        \\
%  $R\sin i$                & 8.68$\pm$0.45 \Rsun                      \\
  Microturbulence         & 1.71$\pm$0.05~\kms                       \\
  Macroturbulence         & 5~\kms (adopted)                         \\
  $[$Fe/H$]$ (solar)       & --0.49$\pm$0.01                          \\
  $T_{\rm eff}$            & 4630$\pm$18 K                          \\
  Luminosity               & 64$\pm$5 \Lsun                           \\
  log$g$ (cgs)             & 2.25$\pm$0.03                            \\
  Radius                   & 12.5$\pm$0.5 \Rsun                       \\
  Inclination              & 44$^{\circ}\pm$10$^{\circ}$          \\
  Mass                     & 1.2$\pm$0.1 \Msun                        \\
  Age                      & $\sim$3.5 Gyr                            \\
%  \noalign{\smallskip}\hline
\hline
 \end{tabular}
\end{center}
\end{table}

\begin{figure}[!htb]
\centering
%\subfloat[Best representation of uncorrected velocities.]
{\includegraphics[angle=0,scale=0.50,clip=true]{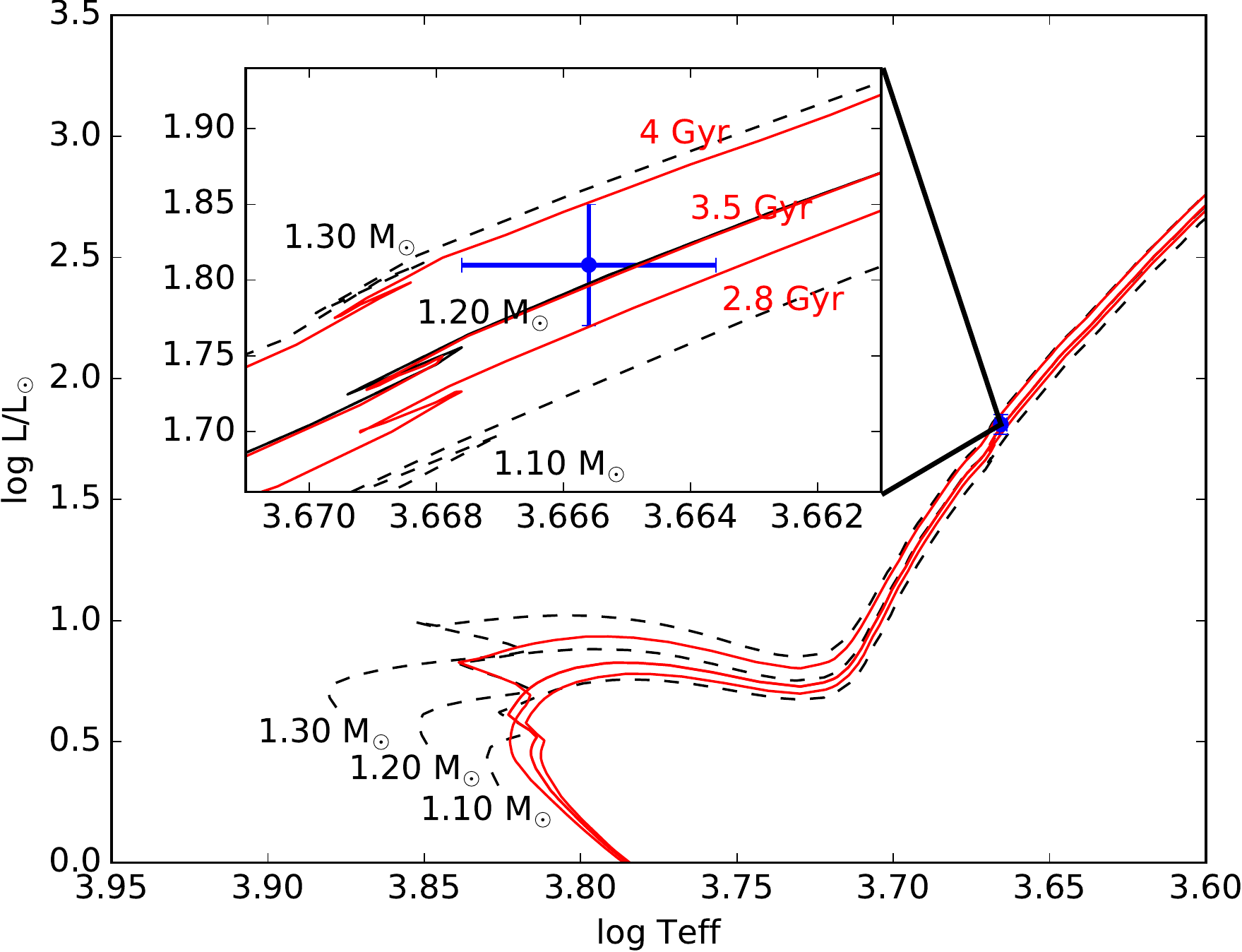}}
%\hfill
%\subfloat[Best representation of spot corrected velocities.]
%{\includegraphics[angle=0,scale=0.50,clip=true]{f4b}}
\caption{Position of HD~208472 in the H-R diagram. The tracks 
(black dashed lines) and isochrones (red continuous lines) are adopted 
from \citet{bert08} for $Y$ = 0.30 and $Z$ = 0.008.}
\label{F4}
\end{figure}

Figure~\ref{F4} compares the position of the star in the $\log T_{\rm eff}$ vs. 
$\log L/\Lsun$ plane with evolutionary tracks and isochrones from \citet{bert08}. 
We chose the $Y$=0.30 and $Z$=0.008 tracks, which are closest to but somewhat less metal 
deficient than our PARSES value (depending on the adopted metallicity of the Sun). 
The inset in Fig.~\ref{F4} provides a detailed zoom into the red-giant branch (RGB). 
The position of HD\,208472 suggests a 1.2 \Msun\ mass and an age of 
$\sim$3.5\,Gyr. It also indicates that the star is past the RGB bump. 
Error ranges of $\sim$0.1\,\Msun\ in mass and an age range of 2.8--4.2\,Gyr are 
estimated from the error bars in Fig.~\ref{F4}.
A summary of the astrophysical parameters is given in Table~\ref{T3}.

Because we do not see evidence of the secondary component at optical wavelengths, 
we expect at least a brightness difference between the components of $\approx$2\fm5. 
In this case, the secondary component must be fainter than $M_V \approx$3\fm2. 
The low-mass function already suggests a low mass for the secondary component. 
With our best-estimate for the inclination and a 1.2\,\Msun \
mass for the primary, 
we estimate $\sim$0.4\Msun\ for the secondary mass. This indicates either a white 
dwarf or a normal M0 dwarf. Unfortunately, there are no ultraviolet observations 
for HD\,208472, and therefore we cannot decide between these two possibilities.

%################ Figure 4 - Evolutionary tracks

\section{Doppler imaging}\label{S4}

\subsection{Technical setup}\label{S4.1}
%iMap code

We used our (Zeeman) Doppler-imaging code $iMap$ \citep{car07,car08,car09,car12} 
to reconstruct the surface temperature distribution from time-series STELLA 
spectra. The code calculates local line profiles by solving the radiative transfer 
equation under LTE assumption and inverts very many 
photospheric absorption lines simultaneously. The iteratively regularized 
Landweber algorithm \citep{engl96,car12} is implemented. During calculations, 
the surface of the star is divided into $5^{\circ}\times 5^{\circ}$ segments; 
2592 segments in total. Local line-profile calculations are made for each segment 
with ATLAS-9 \citep{castelli_kurucz04} model atmospheres for the temperature range 
between 3500\,K and 6000\,K in steps of 250 K and interpolated to the gravity, 
metallicity, and microturbulence value of HD~208472. For further details about the 
code $iMap$, we refer to \citet{car12}.

\subsection{Line selection and data preparation}\label{S4.2}

We selected 34 absorption lines that were relatively blend-free for the given 
temperature and gravity and had a depth larger than 20\%\ of the continuum. 
These lines were taken into account simultaneously during the inversion. 
We applied the multiline technique as described in \citet{car12}. 
Spectral line synthesis was calculated between $\pm$ 50 \kms in the velocity domain, 
however, the velocity range of the line inversion was restricted to $\pm$ 35 \kms 
, where the spectral line wings reach the continuum. The lines and their 
standard parameters from VALD-3  \citep{kupka11} are listed in Table~\ref{T4}.

%################ Table 4 - Chosen lines for DI
\begin{table}[!htb]
\caption{Absorption lines used for the mean profile. Atomic data are taken from VALD-3.}\label{T4}
\begin{flushleft}
 \begin{tabular}{llll}
\hline \hline \noalign{\smallskip}
  Element  &  $\lambda$ (\AA)  &  e.p. (eV) &  $\log(gf)$ \\
\noalign{\smallskip} \hline \noalign{\smallskip}
Fe I  &  5049.8197  & 2.279  &  -1.335  \\
Fe I  &  5307.3607  & 1.608  &  -2.987  \\
Fe I  &  5322.0407  & 2.279  &  -2.803  \\
Cr I  &  5348.3140  & 1.004  &  -1.210  \\
Fe I  &  5367.4659  & 4.415  &  +0.443  \\
Fe I  &  5383.3685  & 4.313  &  +0.645  \\
Fe I  &  5393.1672  & 3.241  &  -0.715  \\
Fe I  &  5415.1989  & 4.387  &  +0.642  \\
Fe I  &  5497.5160  & 1.011  &  -2.849  \\
Fe I  &  5569.6180  & 3.417  &  -0.486  \\
Fe I  &  5572.8423  & 3.397  &  -0.275  \\
Fe I  &  5576.0888  & 3.430  &  -1.000  \\
Ca I  &  5581.9650  & 2.523  &  -0.555  \\
Fe I  &  5862.3564  & 4.549  &  -0.127  \\
Ti I  &  5866.4513  & 1.067  &  -0.790  \\
Fe I  &  5956.6940  & 0.859  &  -4.605  \\
Fe I  &  6021.7889  & 2.198  &  -3.925  \\
Fe I  &  6024.0575  & 4.549  &  -0.120  \\
Ni I  &  6108.1159  & 1.676  &  -2.600  \\
Ca I  &  6122.2170  & 1.886  &  -0.316  \\
Fe I  &  6173.3343  & 2.223  &  -2.880  \\
Fe I  &  6200.3125  & 2.609  &  -2.437  \\
Ti I  &  6261.0988  & 1.430  &  -0.530  \\
Fe I  &  6265.1323  & 2.176  &  -2.550  \\
Fe I  &  6358.6967  & 0.859  &  -4.468  \\
Fe I  &  6393.6004  & 2.433  &  -1.432  \\
Fe I  &  6421.3499  & 2.279  &  -2.027  \\
Fe I  &  6430.8450  & 2.176  &  -2.006  \\
Ca I  &  6439.0750  & 2.526  &  +0.390  \\
Ni I  &  6643.6304  & 1.676  &  -2.220  \\
Fe I  &  6663.4411  & 2.424  &  -2.479  \\
Ca I  &  6717.6810  & 2.709  &  -0.524  \\
Fe I  &  6750.1515  & 2.424  &  -2.621  \\
Ni I  &  6767.7721  & 1.826  &  -2.140  \\
\noalign{\smallskip}\hline
\end{tabular}
\end{flushleft}
\end{table}

We divided the spectra into 13 separate data sets, where each set covers one and a half rotational cycles at most. No set has any spectrum from the 
previous or the following data set in common. This allowed us to trace the spot 
evolution and differential rotation without a biasing correlation between 
subsets that would be due to shared data. We summarize the observing log of all 13 data 
sets in Table~\ref{T5}.

%################ Table 5 - Log of Doppler imaging sets
\begin{table}[!htb]
\caption{Observing log of Doppler-imaging sets.}\label{T5}
\begin{flushleft}
 \begin{tabular}{lcclcc}
\hline \hline \noalign{\smallskip}
Set     &       Year    &       HJD range       &       $N$     &       Max. phase  & rms \\
        &                   &   (2,450,000+)&           &       gap \\
\noalign{\smallskip} \hline \noalign{\smallskip}
1       &       2009.28 &       4923-4944       &       15      &       0.13  & 0.0196  \\
2       &       2009.52 &       5009-5033       &       25      &       0.05  & 0.0021 \\
3       &       2009.59 &       5033-5059       &       25      &       0.07  & 0.0022 \\
4       &       2009.67 &       5060-5089       &       25      &       0.09  & 0.0020 \\
5       &       2010.45 &       5348-5370       &       13      &       0.27  & 0.0019 \\
6       &       2010.53 &       5377-5400       &       14      &       0.14  & 0.0020 \\
7       &       2010.59 &       5401-5419       &       10      &       0.20  & 0.0024 \\
8       &       2010.67 &       5429-5451       &       17      &       0.13  & 0.0019 \\
9       &       2010.73 &       5452-5476       &       10      &       0.31  & 0.0017 \\
10      &       2010.81 &       5478-5507       &       12      &       0.47  & 0.0018 \\
11      &       2011.36 &       5685-5699       &       7       &       0.38  & 0.0018 \\
12      &       2011.47 &       5722-5742       &       17      &       0.13  & 0.0020 \\
13      &       2011.53 &       5743-5763       &       12      &       0.26  & 0.0020 \\
\noalign{\smallskip}\hline
\end{tabular}
\end{flushleft}
\vspace{1mm}

Notes: The year is given for the mid-times of the related sets. $N$ is the number 
spectra per set, and max-phase gap is the largest rotational phase gap between 
two successive observation. The error value of each inverted set is given in the $rms$ 
column.
\end{table}

The rotational phase for the spectra is based on our new orbital elements by using 
the orbital period and a modified epoch as zero point such that one-fourth of 
the orbital period was subtracted from $T_0$ in Table~\ref{T2} (``spot corrected'' column);
\begin{equation}\label{Eq1}
T_0 {\rm (HJD)} = 2,449,246.7642 + 22\fd6194 \ \times \ E .
\end{equation}
At this modified epoch, the giant component is in inferior conjunction, that is, 
zero phase is 0$^{\circ}$ longitude. We assumed that the star rotates in the 
counter-clockwise direction and that the longitude increases in the same direction so that phase 0.25 ($\phi = 0.25$) corresponds to 270$^{\circ}$
longitude on the stellar surface, for instance.

\subsection{$v\sin i$ validation}\label{S4.3}

\begin{figure}[!htb]
\begin{center}
\includegraphics[angle=0,scale=0.40,clip=true]{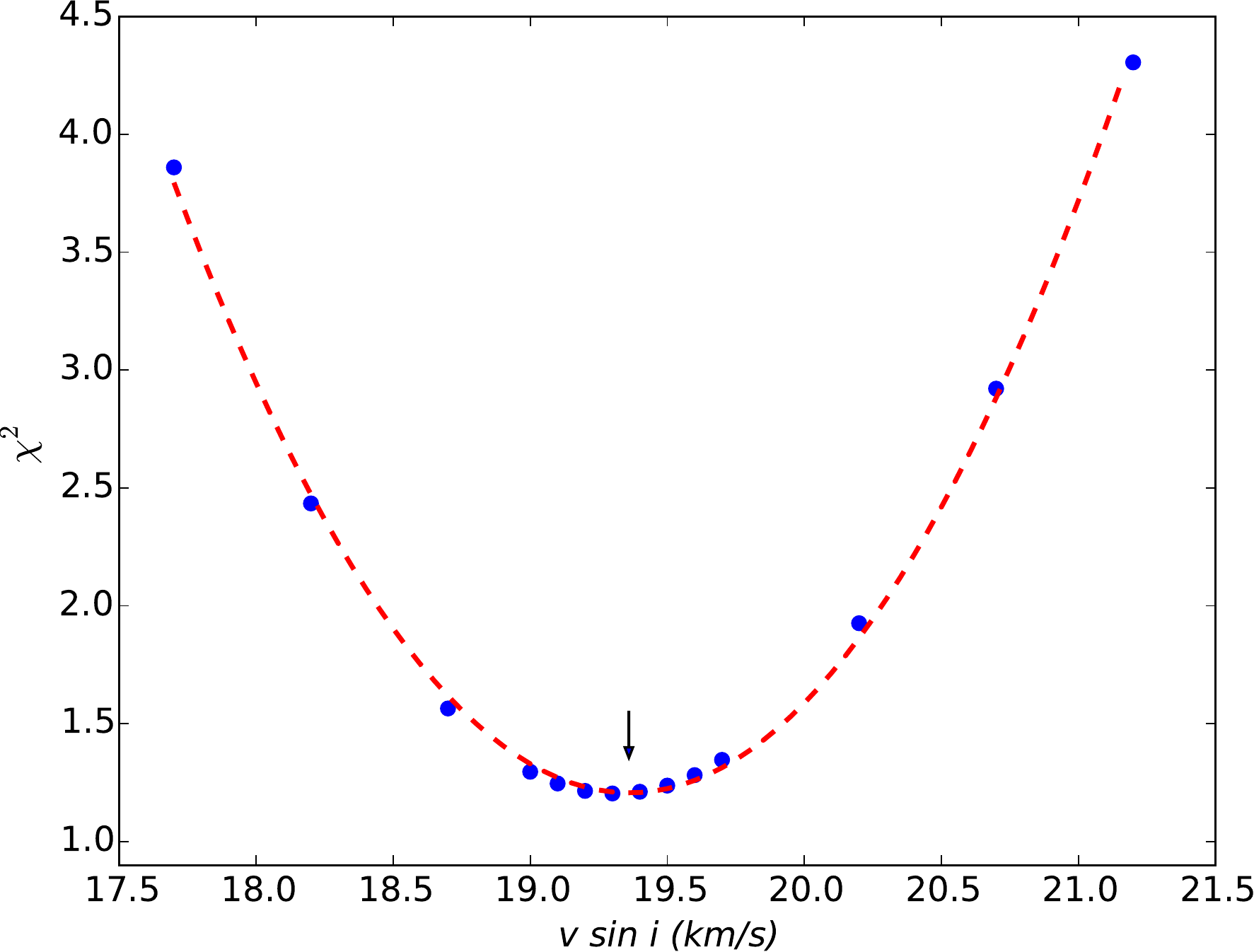}
\caption{Quality-of-fit ($\chi^{2}$) for a series of trial $v\sin i$ values. 
The red dashed curve is a quadratic fit to the data points. The arrow indicates 
the position of the best-fit $v\sin i$ of 19.4\,\kms .}\label{F5}
\end{center}
\end{figure}

Before calculating surface maps, we searched for the $v\sin i$ value that provided 
the best representation for the rotationally broadened line profiles. For this 
purpose, we chose the third data set from the 2009 observing season, which is one 
of the best sets with very good sampling and high S/N. We reconstructed surface images 
for this set by iterating $v\sin i$ values between 17.7 and 21.2~\kms\ while keeping 
the remaining parameters fixed. A parabolic fit to the distribution of 
$\chi^{2}$ in Fig.~\ref{F5} indicates a best-match $v\sin i$ of 19.4$\pm$0.2~\kms.
The error comes from the width of the parabola, $\pm$0.2~\kms steps in the 
sampling, where the deviation is 1$\sigma$. We adopted this value as our best estimate 
and kept it fixed for the remaining imaging.

%################ Figure 5 - vsini search

\section{Results}\label{S5}

\subsection{Surface maps}\label{S5.1}

\begin{figure*}[!htb]
\begin{center}
\begin{tabular}{cc}
{\bf Set 1} & \raisebox{-.5\height}{\includegraphics[scale=0.3]{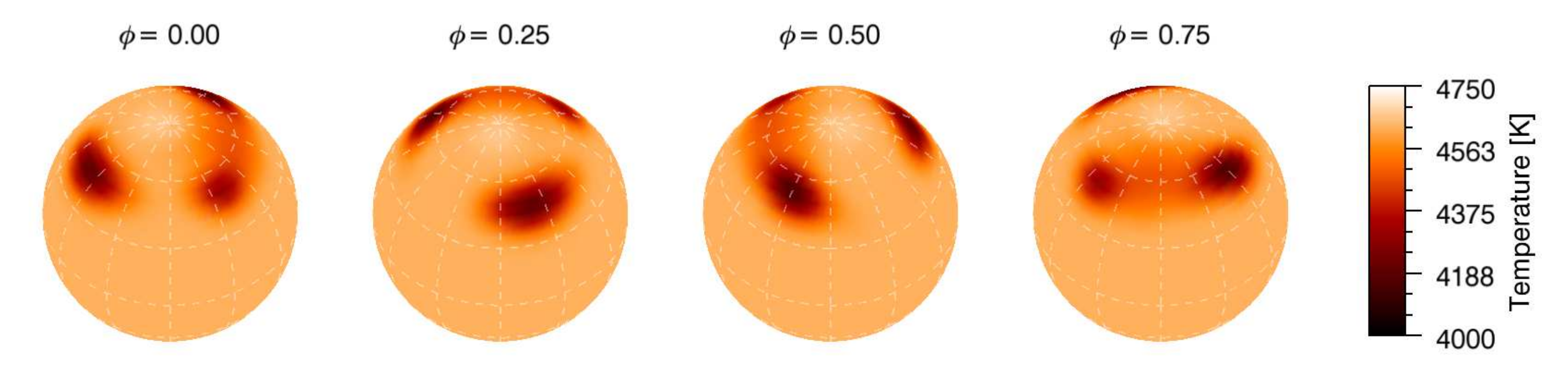}}\\
{\bf Set 2} & \raisebox{-.5\height}{\includegraphics[scale=0.3]{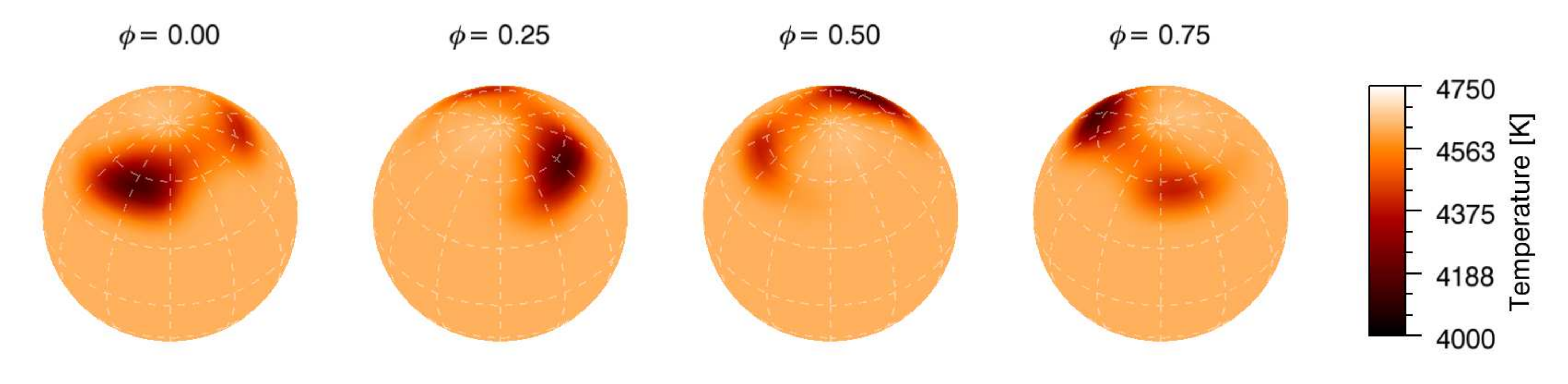}}\\
{\bf Set 3} & \raisebox{-.5\height}{\includegraphics[scale=0.3]{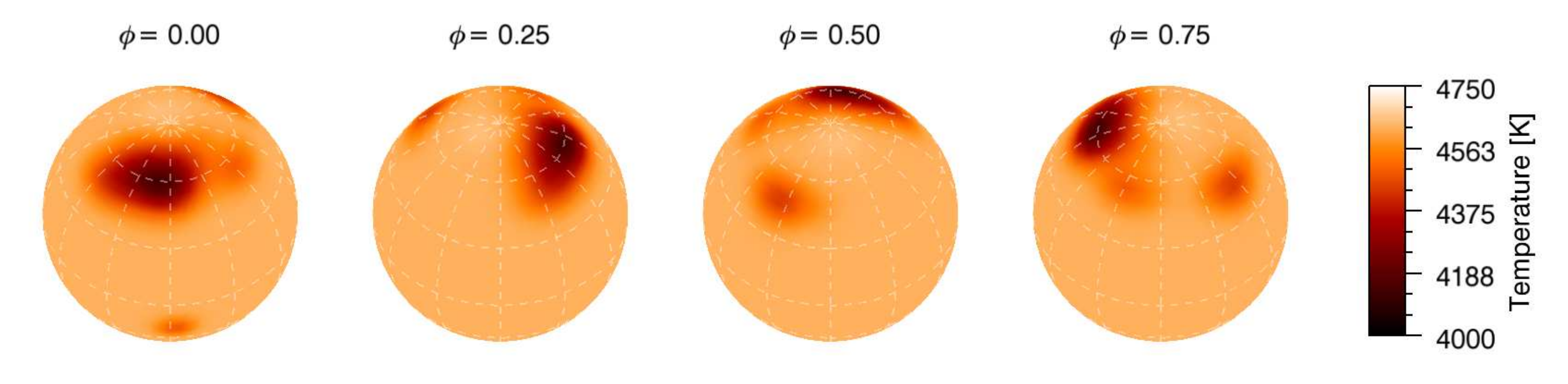}}\\
{\bf Set 4} & \raisebox{-.5\height}{\includegraphics[scale=0.3]{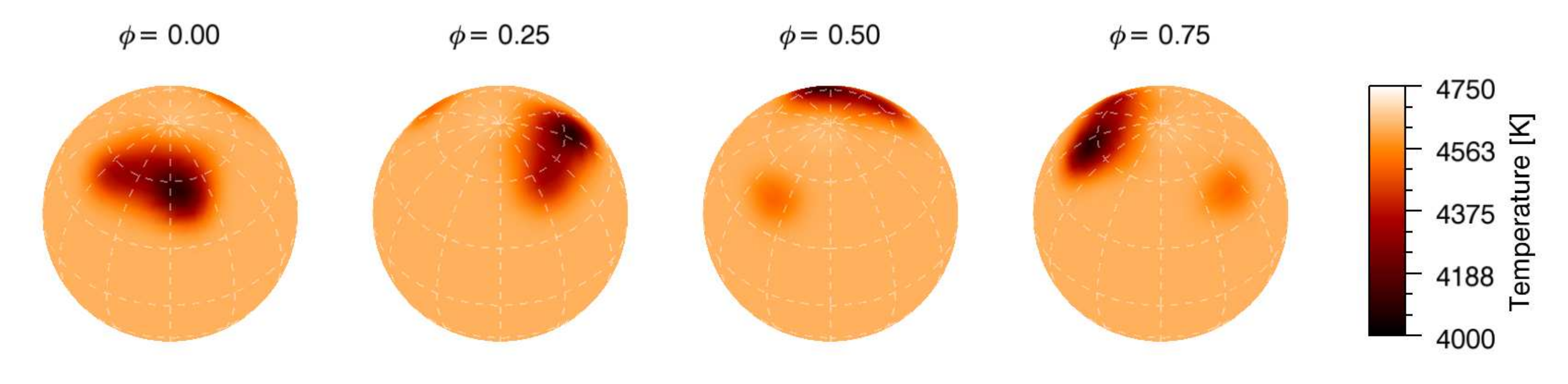}}\\
\end{tabular}
\end{center}
\caption{Doppler images of HD~208472 for the 2009 observing season.}\label{F6}
\end{figure*}

\begin{figure*}[!htb]
\begin{center}
\begin{tabular}{cc}
{\bf Set 5} & \raisebox{-.5\height}{\includegraphics[scale=0.3]{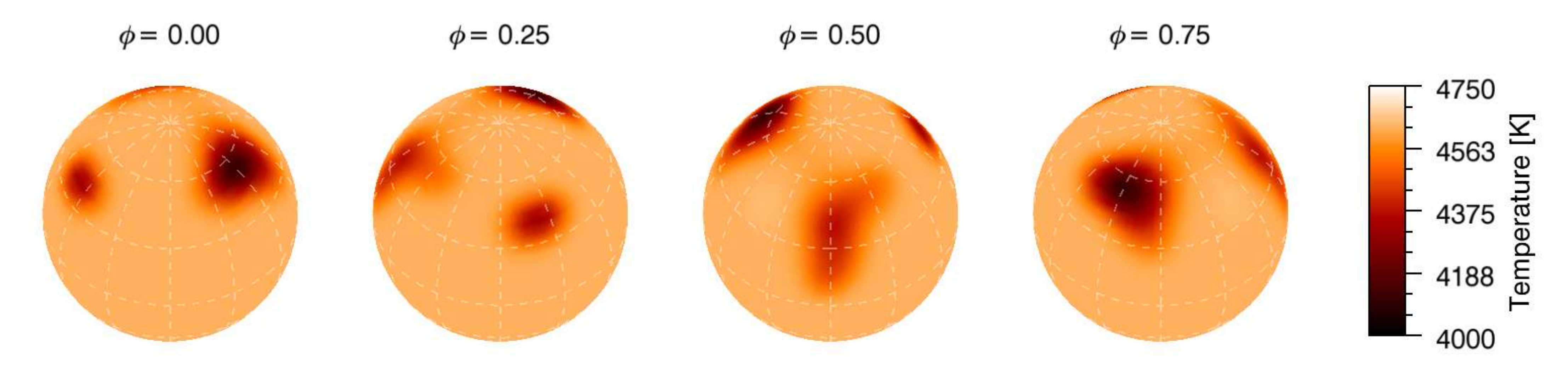}}\\
{\bf Set 6} & \raisebox{-.5\height}{\includegraphics[scale=0.3]{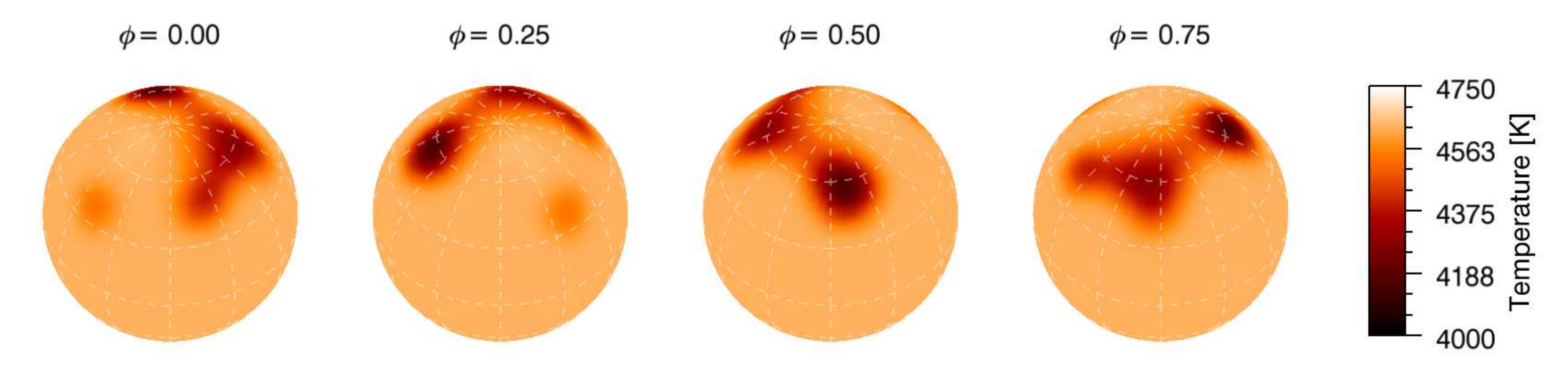}}\\
{\bf Set 7} & \raisebox{-.5\height}{\includegraphics[scale=0.3]{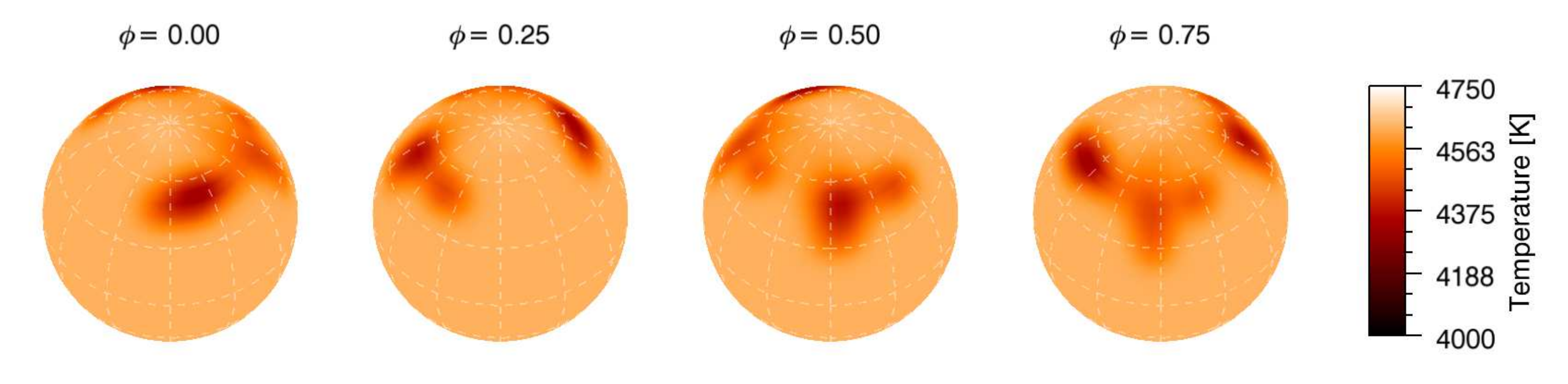}}\\
{\bf Set 8} & \raisebox{-.5\height}{\includegraphics[scale=0.3]{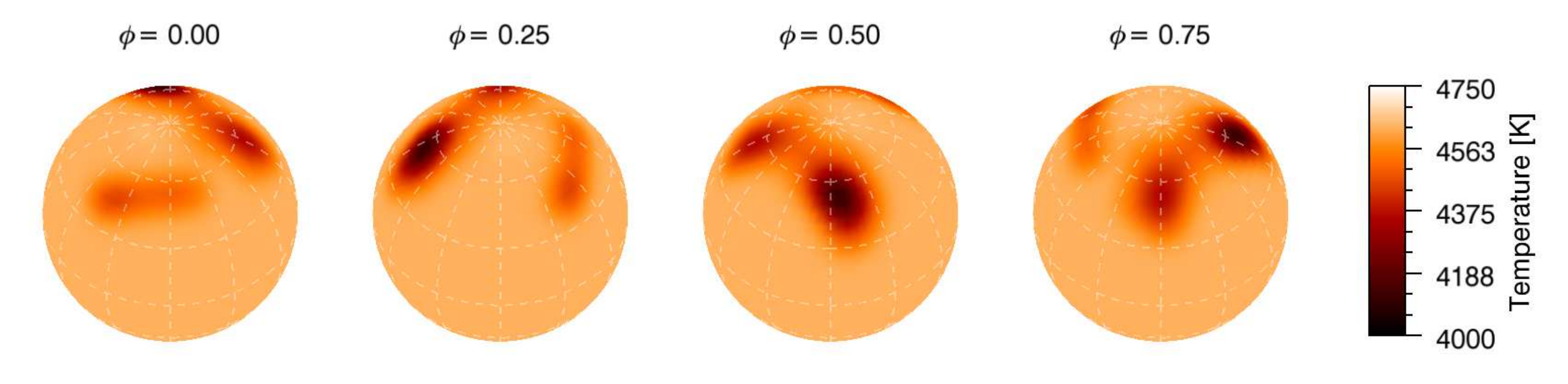}}\\
{\bf Set 9} & \raisebox{-.5\height}{\includegraphics[scale=0.3]{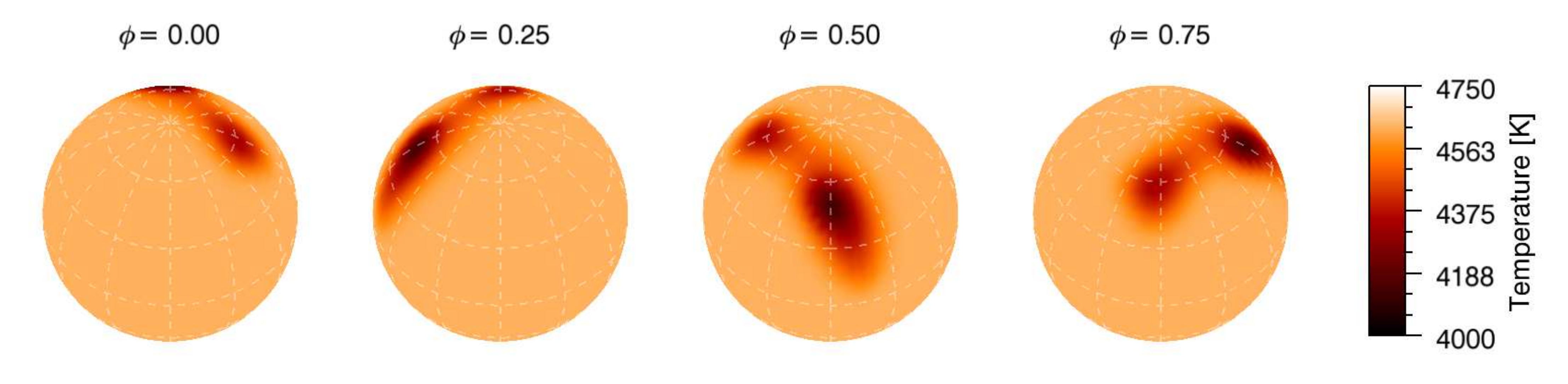}}\\
{\bf Set 10} & \raisebox{-.5\height}{\includegraphics[scale=0.3]{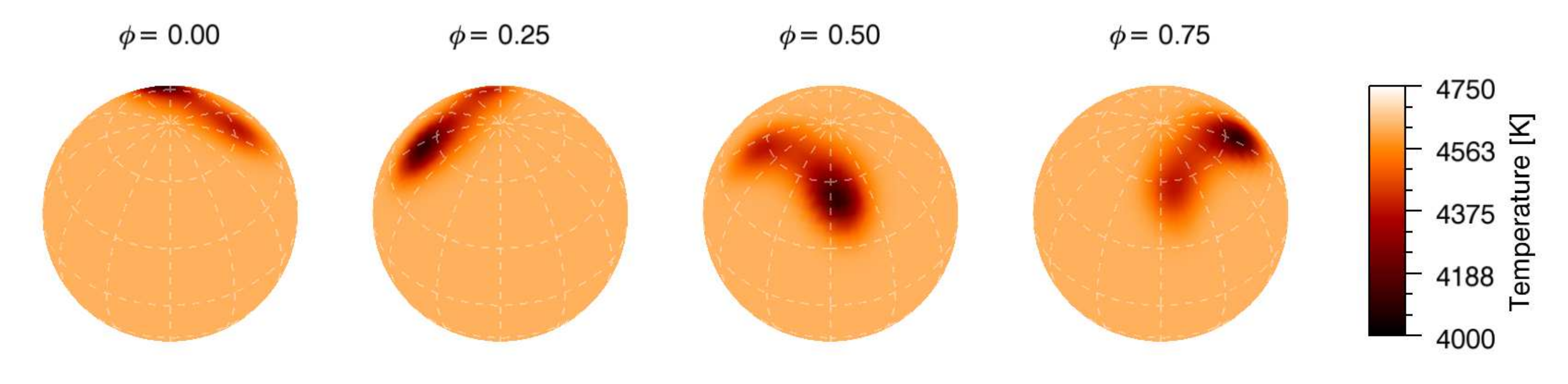}}\\
\end{tabular}
\end{center}
\caption{Doppler images of HD~208472 for the 2010 observing season.}\label{F7}
\end{figure*}

\begin{figure*}[!htb]
\begin{center}
\begin{tabular}{cc}
{\bf Set 11} & \raisebox{-.5\height}{\includegraphics[scale=0.3]{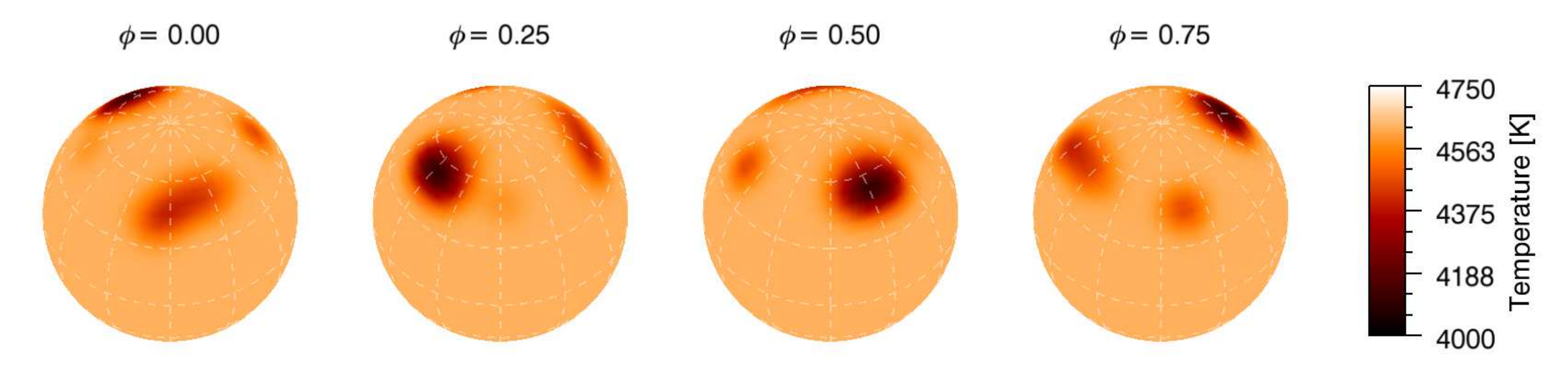}}\\
{\bf Set 12} & \raisebox{-.5\height}{\includegraphics[scale=0.3]{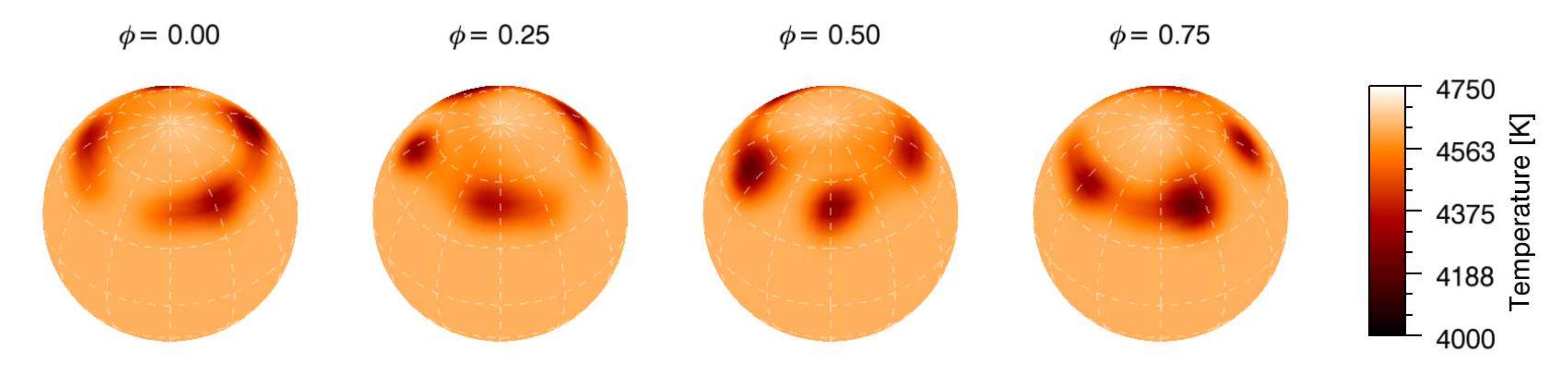}}\\
{\bf Set 13} & \raisebox{-.5\height}{\includegraphics[scale=0.3]{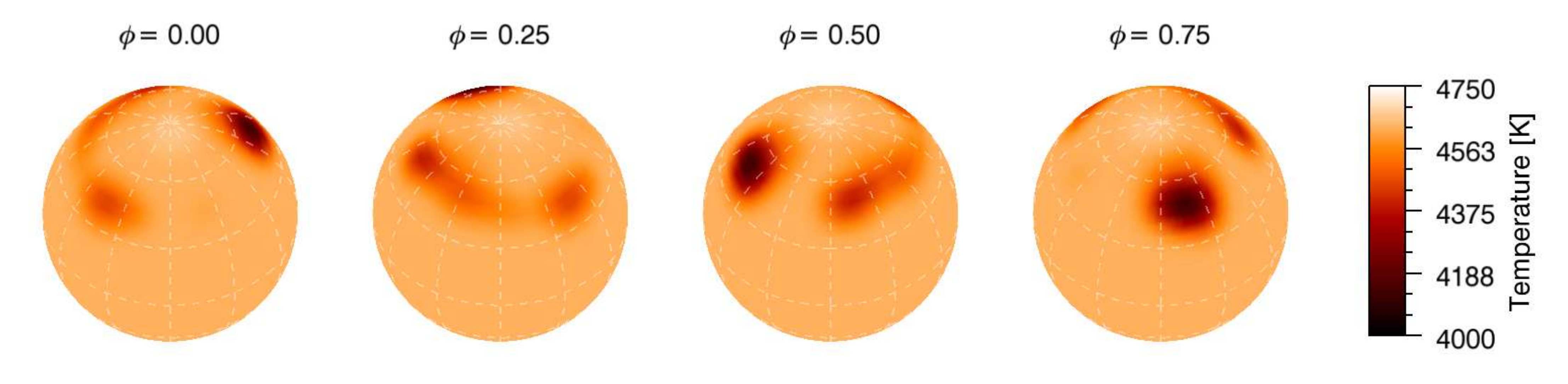}}\\
\end{tabular}
\end{center}
\caption{Doppler images of HD~208472 for the 2011 observing season.}\label{F8}
\end{figure*}

Figures~\ref{F6} to \ref{F8} present our Doppler images for 
2009 to 2011. We plot observed and inverted line profiles
of each set in Fig.~\ref{A1}. None of the images shows a dominant polar 
spot, in agreement with earlier maps by \citet{weber04}. Spots on HD\,208472 
are generally not very large and on average seem cooler than the undisturbed 
photosphere by just $\sim$500\,K. We did not recover any bright features warmer 
than the photosphere.

%################ Figure 6 - Doppler maps 2009

%################ Figure 7 - Doppler maps 2010

%################ Figure 8 - Doppler maps 2011

%################ Figure 9 - photometric spot longitudes

\begin{figure*}[!htb]
\vspace{0.5cm}
\begin{tabular}{cc}
{\bf a)} & \raisebox{-0.9\height}{\includegraphics[angle=0]{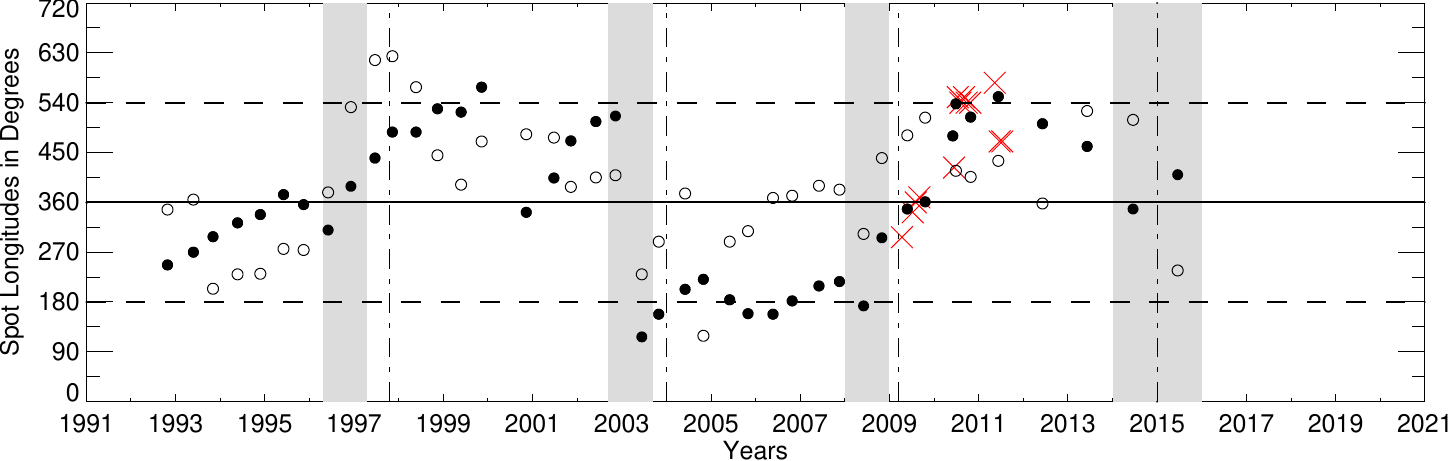}}\vspace{0.5cm}\\
{\bf b)} & \raisebox{-0.9\height}{\includegraphics[angle=0]{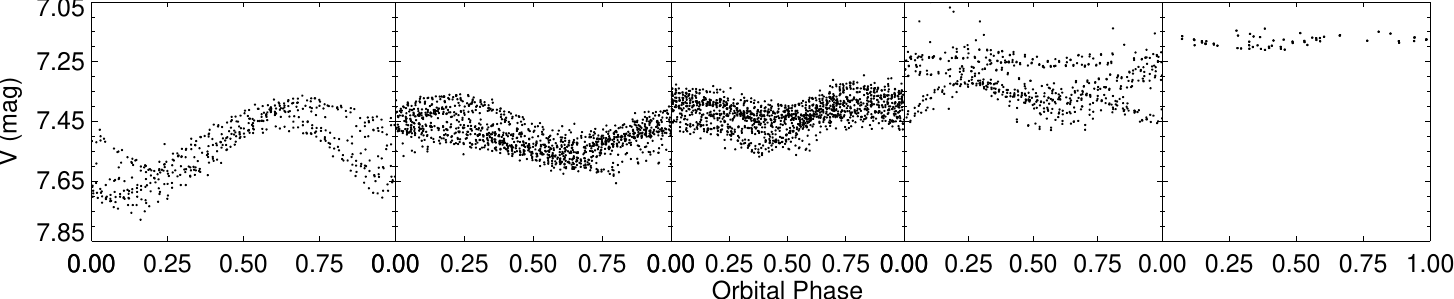}}\\
%{\bf a)} \\
%\hspace*{1cm}\includegraphics[angle=0]{f9a}\\
%{\bf b)} \\
%\hspace*{0.92cm}\includegraphics[angle=0]{f9b}\\
\end{tabular}
\caption{\textbf{a.} Longitudinal distribution of spots on HD~208472 from 
photometric two-spot models. The shaded regions are the time ranges where 
spots change their preferential positions (hemispheres) 
on the surface of the star. The continuous line represents the 
longitudes of the apsidal line of the binary system, while the two dashed 
lines restrict the stellar surface. The vertical dot-dashed lines 
correspond to times when changes occurred in light-curve amplitude and 
photometric period. We overplot the longitudes of the main spot 
of the Doppler images with cross symbols in red. 
\textbf{b.} Accumulated $V$ brightness phased with 
the ephemeris in Eq.~\ref{Eq1}, and separated into the corresponding time 
intervals in panel~a.}\label{F9}
\end{figure*}

Spots typically appear at latitudes higher than $\approx$30$^{\circ}$, 
except for a few cases where an extension of a larger spot may bridge 
that latitude (e.g., in 2010, sets~5 and~9). The  longitudinal concentration, 
on the other hand, had changed abruptly at one point. While in 2009, spots 
concentrated around zero phase ($\phi = 0.00$; facing the observer at inferior 
conjunction), in the following season in 2010 the main concentration occurred 
near $\phi = 0.50$ and remained so into 2011 until the end of our data set. 
We focus on this change in Sect.~\ref{S5.2}, but first we describe the results 
from the individual observing seasons.

\emph{2009 observing season.} We have complete phase coverage for each image 
reconstructed from 2009. In the first image (set~1), we reconstruct three 
individual spots with latitudes between 45$^{\circ}$ and 75$^{\circ}$ and 
separated by $\sim$0.40 in phase. The second image (set~2) shows the larger 
of the three spots at $\sim 60^{\circ}$ latitude and at $\phi = 0.05$. Two of 
the spots seem to be marginally connected with each other, and this region 
appears to evolve systematically. Because there are almost four rotations 
between sets~1 and~2, some spot evolution is even expected. As time 
progresses from set~2 to set~4 (i.e., until the last image in 2009), the main spot at 
$\phi = 0.05$ still exists but appears to have drifted toward decreasing 
phases with respect to the epoch and orbital period in Eq.~\ref{Eq1}. 
 Spot migration like this is also typical for the shifts expected from differential 
rotation. One of the two originally connected spots had continuously drifted 
toward the larger spot.  It had completely disappeared by the end of the season.

\emph{2010 observing season.} In 2010 the phase coverage is not as good as 
in 2009, but still quite good for reconstructing reliable surface maps. The time 
gap between successive maps is typically a few  days, except between sets~7 and~8, 
where it is ten days. Nevertheless, spot evolution can be traced safely through 
six successive stellar rotations. In the beginning of 2010 (set~5), we see the 
main spot at $\phi = 0.83$ and two weaker spots at $\phi = 0.50$ and 
$\phi = 0.18$ at lower latitudes than the main spot. As time progresses, 
the spot at $\phi = 0.18$ fades, while the spot at $\phi = 0.50$  cools and 
becomes the most dominant feature. The spot at $\phi = 0.83$ still survives but 
 warms up and eventually approaches the main spot at $\phi = 0.50$.
We also observe that the two spots are connected to each other with a pale structure 
around $\sim 75^{\circ}$ latitude that is slightly cooler than the photosphere. All spots appear paler in set~7, while these spots appeared darker in the 
previous and in the following sets. This is most likely related to the data quality 
of set~7, where the average S/N is just around 120, while it is usually between 
170--200. Moreover, the visible surface from sets 9 and 10 at 
$\phi = 0.00$ appears to be almost unspotted except for the high-latitude features close to the 
rim of the visible surface. We emphasize that sets~9 and~10 have considerable phase 
gaps (listed in Table~\ref{T5}). We have no observation between $\phi = 0.75$ and 
$\phi = 0.25$ in set~10 and $\phi = 0.70$ and $\phi = 0.00$ in set~9. The consequence 
is that whatever weak structure may have existed at these longitudes, it cannot be 
reconstructed. The sudden disappearance of spots within these phases is 
related to these phase gaps and not to spot decay, for example.

\emph{2011 observing season.} We have three maps for 2011 covering four consecutive 
stellar rotations. There is a time gap of one rotation between the end of set~11 and 
the beginning of set~12. We note that set~11 comprises only seven spectra with a 
maximum phase gap between two successive observations of 0.38 between $\phi = 0.28$ and 
$\phi = 0.66$. Care must be taken to avoid overinterpreting spots at these phases. 
Set~11 shows one cool, dominant spot at $\phi = 0.40$ and two warmer 
spots at $\phi = 0.70$ and $\phi = 0.00$. After two rotations (i.e., in set~12), the main spot appears to have broken into two smaller and warmer spots, while the spot at 
$\phi = 0.70$ became the most dominant spot. However, this evolution scenario is hampered 
by the phase gaps in set~11 and is not conclusive. The spot at $\phi = 0.00$ seems to 
have moved toward decreased phases by the time of set~12 and then has vanished in 
set~13. At the same time, a small appendage of the originally cooler spot has migrated 
to $\phi = 0.10$.

\subsection{Active longitudes}\label{S5.2}

In paper~I we showed that an activity cycle repeats almost regularly in time. 
A preliminary 6.28-year activity cycle was reconstructed from long-term photometry. 
It predicted a new activity cycle to begin around 2009/2010. In the previous section, 
we noted that the main spot activity was located around $\phi = 0.00$ in 2009, while 
it was around $\phi = 0.50$ in 2010 and 2011. There were also
almost 
no spots at $\phi = 0.50$ in the 2009 maps, while the main spot concentration was 
observed at exactly that phase in 2010 and 2011. This means that we apparently witnessed 
a flip-flop event. The new longitudes of the main activity zone appear to correspond 
to either the sub-stellar point (i.e., $\phi = 0.00$) or its antipode ($\phi = 0.50$). 
We interpret this as evidence for two persistent active longitudes separated by 180$\,^{\circ}$ 
on the surface of HD\,208472, which might be related with the binary nature of 
the star.

We also applied the same two-spot model technique as in paper~I to our recent $V$-band 
photometry from 2010--2015. In this way, we obtained the longitudinal distribution 
of spots, but over a longer period in time than the Doppler imaging. For details of the 
technique we refer to paper~I and references therein. The results are shown 
in Fig.~\ref{F9}a, which is the updated version of the Fig.~\ref{F10} in paper~I. 
The spot positions show the continuing systematic drifts and verify the beginning of 
the new activity cycle around 2009/2010. Furthermore, the most recent photometric data 
show some evidence for yet another phase jump in 2015, in agreement with the 
prediction from the  $\sim$6 yr cycle length. The light curves in the orbital phase 
frame are shown in Fig.~\ref{F9}b. The active longitudes clearly show the 
alternating locations of minima, but the figure also emphasizes the existence of the ever 
increasing brightness of the system (see Fig.~\ref{F1}).

\subsection{Differential rotation}\label{S5.3}

We traced the position of spots by cross-correlating successive Doppler maps in 
longitudinal direction and examining the amount of longitudinal shift in a given 
latitude range where spots occur (e.g., \citealt{kov15,kunstler15}). This is basically 
equivalent to tracking the motion of sunspots as applied to solar-disk images 
(e.g., \citealt{how84,brajsa02}).

\begin{figure}[!htb]
\begin{flushleft}
{\bf a)} \\
\includegraphics[angle=0,scale=0.68,clip=true]{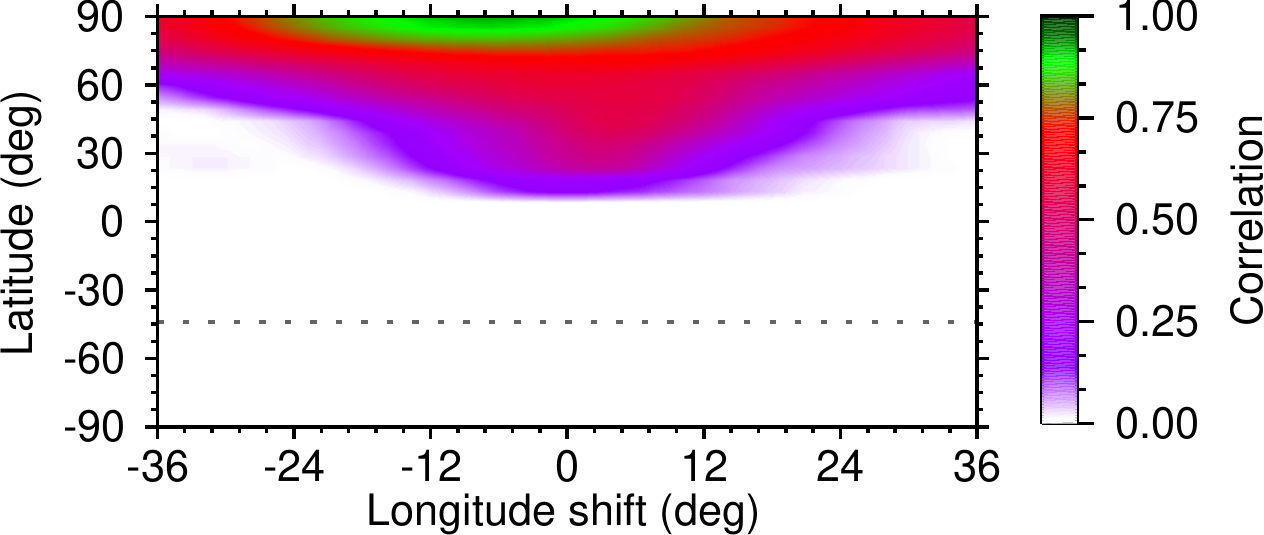}\\
{\bf b)} \\
\includegraphics[angle=0,scale=0.5,clip=true]{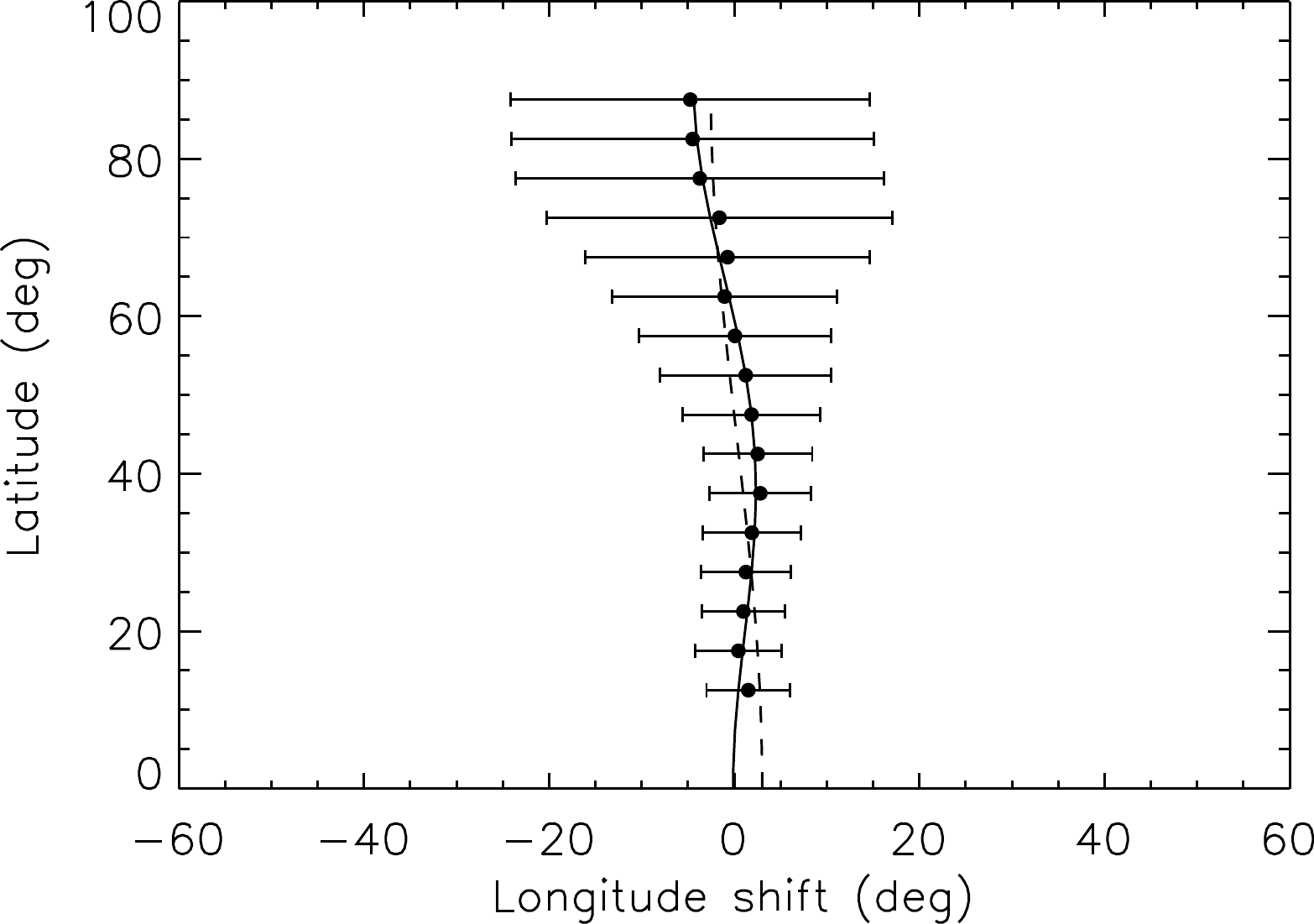}
\end{flushleft}
\caption{\textbf{a)} Average ccf map from ten seasonal ccf maps.
\textbf{b)} Average differential rotation pattern. The dashed line denotes the best 
fit using the single-term law in Eq.~\ref{Eq2}, while the solid line shows the best 
fit from the double-term law in Eq.~\ref{Eq3}. }
\label{F10}
\end{figure}

In our present application, we are able to produce three cross-correlation-function 
(ccf) maps for the observing season 2009, five ccf maps for 2010, and two ccf maps 
for 2011. Altogether, we have ten ccf maps at our disposal. Because the Doppler images 
are based on a 5$^{\circ}\times 5^{\circ}$ equal-degree segmentation of the surface, 
we also calculated the cross correlation for latitudinal bins of 5$^{\circ}$ width. 
A Gaussian profile was used for fitting and identifying the CCF peaks. We present the 
grand average ccf map in Fig.~\ref{F10}a, which was obtained by averaging all ten 
ccf maps from three consecutive years. It shows that the available spot-displacement 
information extends from the visible pole down to $\approx$10$^{\circ}$ latitude. 
Thus, \object{HD\,208472} provides a wide latitude range with spots and indeed allows 
for a good estimation of its differential rotation. In Fig.~\ref{F10}b we plot the 
positions of the ccf peaks along with their estimated errors defined as the full-width 
at half maximum of the respective Gaussian fits. We then fit a solar-like differential rotation 
law to the global ccf pattern of the star. Considering the angular velocities at a 
latitude $b$ and at the equator ($\Omega(b)$ and $\Omega_{eq}$, respectively), and the 
difference between the angular velocities at the pole and the equator 
($\Delta\Omega = \Omega_{\rm eq} - \Omega_{\rm pole}$), we adopt a typical differential 
rotation law as
\begin{equation}\label{Eq2}
\Omega(b) = \Omega_{eq} - \Delta\Omega sin^{2}(b) \ ,
\end{equation}
where the surface shear parameter, $\alpha$, is defined by 
$\alpha$ = $\Delta\Omega / \Omega_{\rm eq}$. Using Eq.~\ref{Eq2} to fit the 
differential rotation pattern in Fig.~\ref{F10}b, we find $\alpha$ = +0.015$\pm$0.003 and 
$\Omega_{\rm eq}$ = 0.2801 $\pm$ 0.0004 rad/day, which corresponds to an equatorial period 
$P_{\rm eq}$ = 22.431$\pm$0.031\,d.

A better representation of the differential rotation pattern can be achieved by adding 
another term to the differential rotation law, which is commonly adopted for the Sun,
\begin{equation}\label{Eq3}
\Omega(b) = \Omega_{\rm eq} + \Omega_{1} sin^{2}(b) + \Omega_{2} sin^{4}(b)
,\end{equation}
where $\Omega_{\rm pole}$ is given by the sum of $\Omega_{\rm eq}$, $\Omega_{1}$, and 
$\Omega_{2}$. In this case, we find a shear parameter $\alpha$ of +0.012$\pm$0.008, 
$\Omega_{\rm eq}$ = 0.2777$\pm$0.0006 rad/day and $P_{\rm eq}$ = 22.628$\pm$0.048\,d,
where $\Omega_{1}$ = 0.0101$\pm$0.0016 rad/day and $\Omega_{2}$ = --0.0134$\pm$0.0015 rad/day.

Both obtained values of $P_{\rm eq}$ differ by more than their individual errors. 
This discrepancy is caused by the lack of data points below a latitude of $\pm$10$^{\circ}$ deg. 
The two $\alpha$ values agree to within their errors and suggest weak but solar-type differential 
rotation with a surface shear about one order of magnitude weaker than for the Sun. Such weak 
differential rotation is a common result for RS~CVn binary stars with late-type cool giants, 
such as for XX~Tri \citep{kunstler15} or $\zeta$~And \citep{kov12}, for which both studies were based 
on time-series Doppler imaging.

%################ Figure 10 - ccf map and dif. rot.

\section{Summary and conclusions}

Three years of high-resolution time-series STELLA spectra enabled us to not only 
refine precise spectroscopic orbital elements and atmospheric parameters of HD~208472, 
but reconstruct 13 separate surface images from three observing seasons through 
multiple-line inversions. Surface reconstructions were made by tracing the distortions 
in the absorption line profiles with time, which is the main principle of our 
time-series approach of Doppler imaging. The line-profile distortions are also the 
cause of systematic shifts in radial velocities, known as spot jitter in case of 
chromospherically active stars. After proper removal, it decreased the RV residuals 
for our new spectroscopic orbital solution by a factor of four from 380~\ms\ to 88~\ms. 
Fourier analysis of the RV residuals indicated that the dominant periods are always 
very close to the orbital period and its harmonics, which is the typical signature 
of spot jitter.

HD~208472 exhibits active longitudes on adjacent hemispheres and occasional flip-flops 
between them that overlap the continuous drift in longitude with respect to the orbital 
reference frame. Similar active longitudes and flip-flops were observed in several 
RS~CVn binaries \citep{jetsu96,berd_tuo98,kunstler15}, FK~Com-type stars 
\citep{jetsu91,jetsu99}, and young solar-like stars \citep{berd05}. There is some 
evidence that flip-flops occur more or less regularly with timescales of a few years 
up to a decade, sometimes referred to as flip-flop cycle (see the references above), 
but the evidence remains inconclusive. To detect these cycle lengths, we still 
need long-term photometry with phase-resolved sampling or, even better, modern time-series 
Doppler imaging.

A 6.28 yr flip-flop cycle was discovered for HD~208472 in our paper~I, where we 
predicted the begin of a new cycle for 2009/2010. Our Doppler imagery in the present 
paper caught this event. We witnessed the change of the main activity zone from 
$\phi = 0.00$ to $\phi = 0.50$ between the 2009 and 2010 observing seasons over 
a time range of approximately one year. Thus, the location of the main activity 
zone corresponds to the sub-stellar point and its antipode, as
has been suggested by 
\citet{olah06}. Theoretically, coexistence of axisymmetric and non-axisymmetric dynamo 
modes may produce active longitudes and also give rise to the flip-flop  phenomenon 
\citep{fluri04,moss05,elstner05}. However, why these are aligned and influenced by 
the secondary star and why the brightness continues to increase over several 
activity cycle is yet to be understood.

From cross-correlating successive Doppler maps, we found a surface shear of 
$\alpha$ = +0.015$\pm$0.003 for a $\sin^2$-latitude differential rotation law and 
$\alpha$ = +0.012$\pm$0.008 for a $\sin^2 + \sin^4$-latitude differential rotation law. 
This differential rotation is in solar-like direction, meaning
that poles rotate slower than 
the equator, but $\approx$ 15 times weaker than on the Sun. The average value is close 
to the $\alpha$ value of \object{XX~Tri} \citep{kunstler15}, which is a similar RS~CVn 
system in terms of physical properties and spectroscopic orbit, but maybe$\text{ }$\text{about twice as} 
old as HD~208472. Comparing other stars whose $\alpha$ values were determined from 
Doppler imaging, our target star belongs to a group of weak solar-type differential rotators 
(\object{AB~Dor}, \citealt{donati97}; \object{$\zeta$~And}, \citealt{kov12}; \object{IL~Hya}, 
\citealt{kov14}). On the other hand, anti-solar differential rotation was found 
(or at least claimed) in some other active stars, such as \object{$\sigma$~Gem} 
\citep{kov15}, \object{UZ~Lib} \citep{vida07}, \object{HU~Vir} \citep{str94,gohar}, and 
\object{HD~31993} \citep{strass03}. Tidal force in a binary system is considered as a 
possible explanation for the existence of two different directions of differential 
rotation, depending on the orbital and physical properties of the related system. 
\citet{holz_02} showed that the tidal force in an RS~CVn binary may lead to preferred 
longitudes. It may also affect the angular momentum distribution in the convective 
envelope of the giant component and alter its differential rotation pattern \citep{kov12}. 
Another effect of the tidal force on differential rotation might be the suppression of the 
strength of the differential rotation through tidal locking \citep{col_cam07}. 
The RS~CVn binaries mentioned above have $\alpha$ values of just a few of per cent 
(solar or anti-solar). In comparison, the single K2 giant \object{HD~31993} \citep{strass03}, 
where no tidal interaction is present, has a dramatically higher shear value of 
$\alpha = -0.125$. Another comparison can be made with the single K1 giant \object{DP~CVn}
\citep{kov13}, which has a surface shear value of $\alpha = -0.035$, which is a stronger surface shear 
in magnitude than \object{HD~208472}. This shows that  fast-rotating single giants
with $\alpha$ values even comparable to the solar value have been reported, while no such
RS CVn stars have been found. This supports the picture that differential rotation is
suppressed by tidal forces as suggested by \citet{schar82}, for
example.

\begin{acknowledgements}

O\"O thanks the Scientific and Technological Research Council of Turkey (T\"UB\.ITAK) 
for funding this study under ``2219 post-doc research scholarship program''. KGS thanks the 
State of Brandenburg and the German BMBF for the continuous funding of the STELLA facility in 
Tenerife and the APTs in southern Arizona.

\end{acknowledgements}


\begin{appendix}
\section{Line profiles of Doppler images}

%################ Figure A1 - Observed and inverted line profiles
\begin{figure*}[!htb]
\begin{center}
\begin{tabular}{cccccc}
{\small 2009 - set 1} & {\small 2009 - set 2} & {\small 2009 - set 3} & {\small 2009 - set 4} & 
{\small 2010 - set 5} \\
\includegraphics[scale=0.185,clip=true]{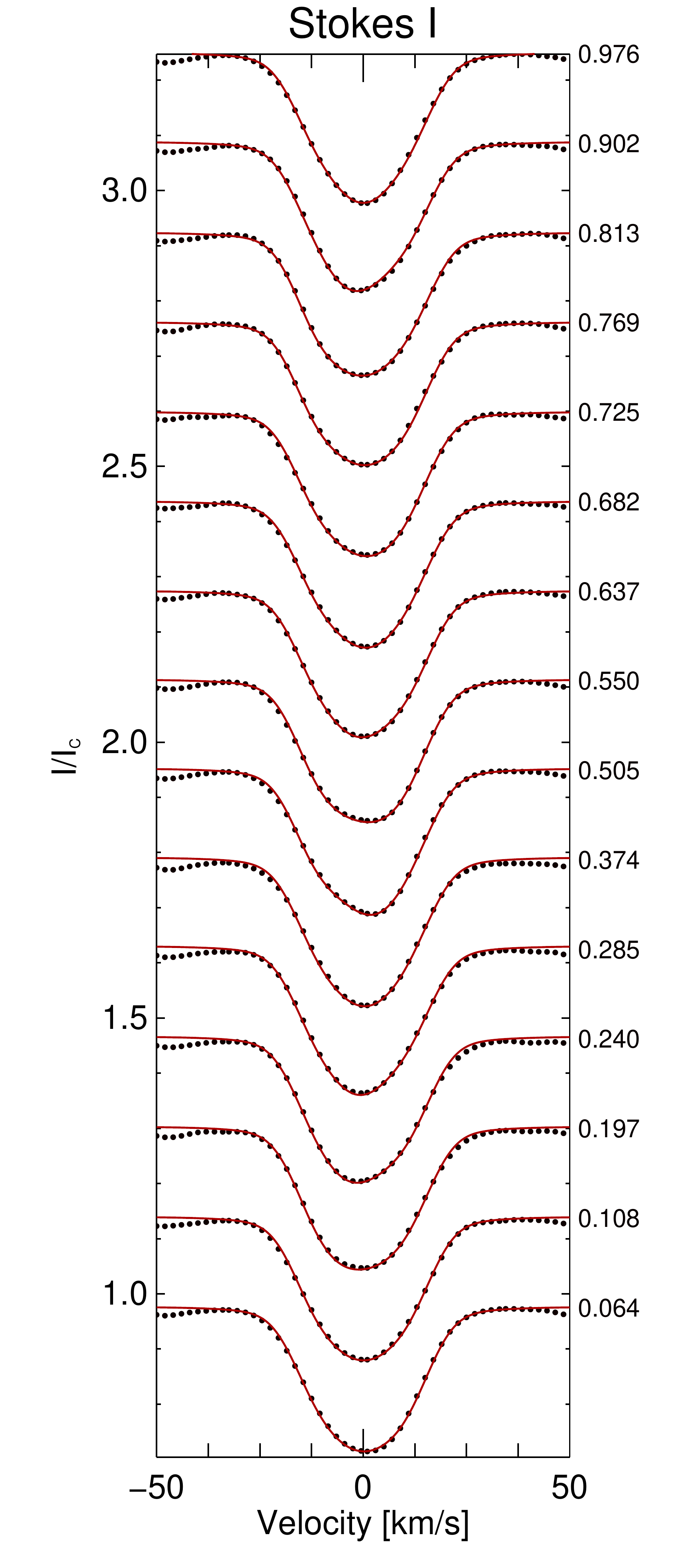}&
\hspace{-0.5cm} \includegraphics[scale=0.185,clip=true]{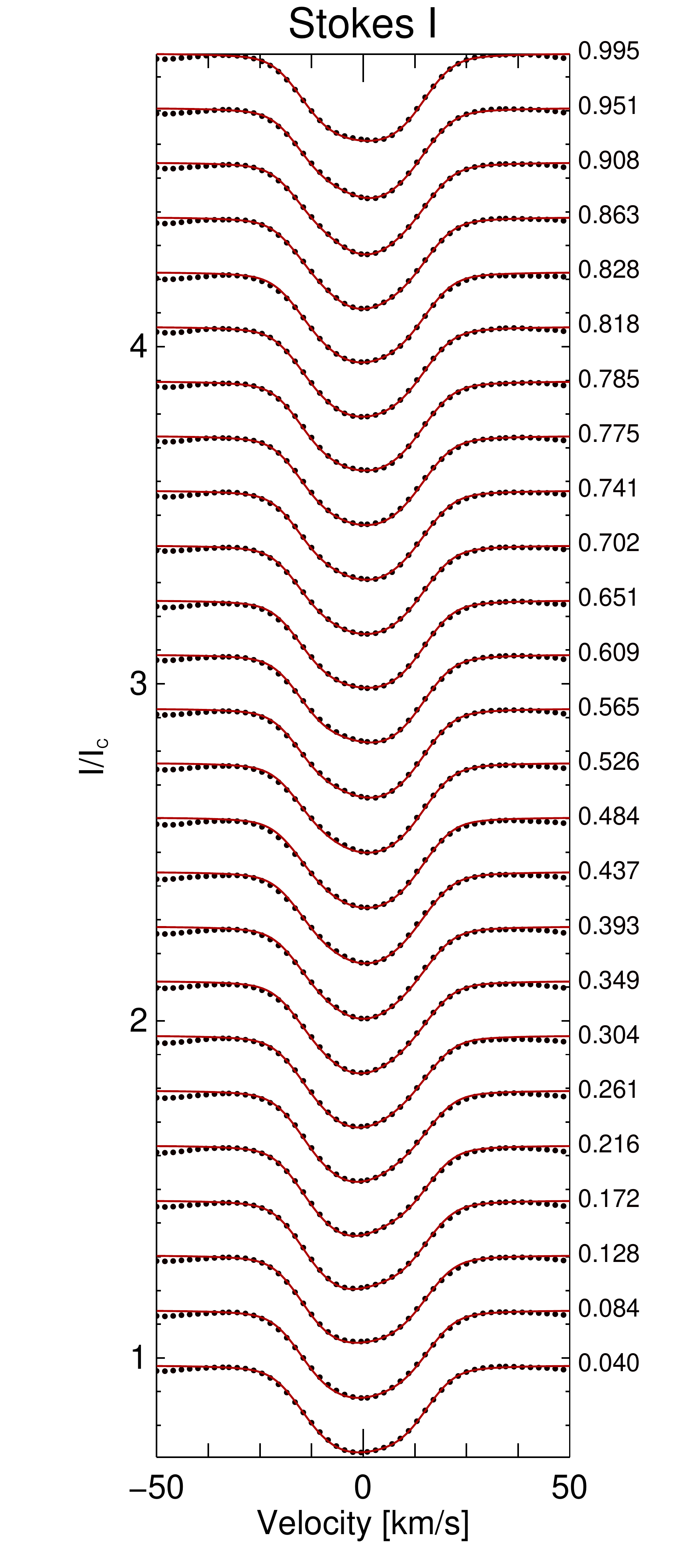}&
\hspace{-0.5cm} \includegraphics[scale=0.185,clip=true]{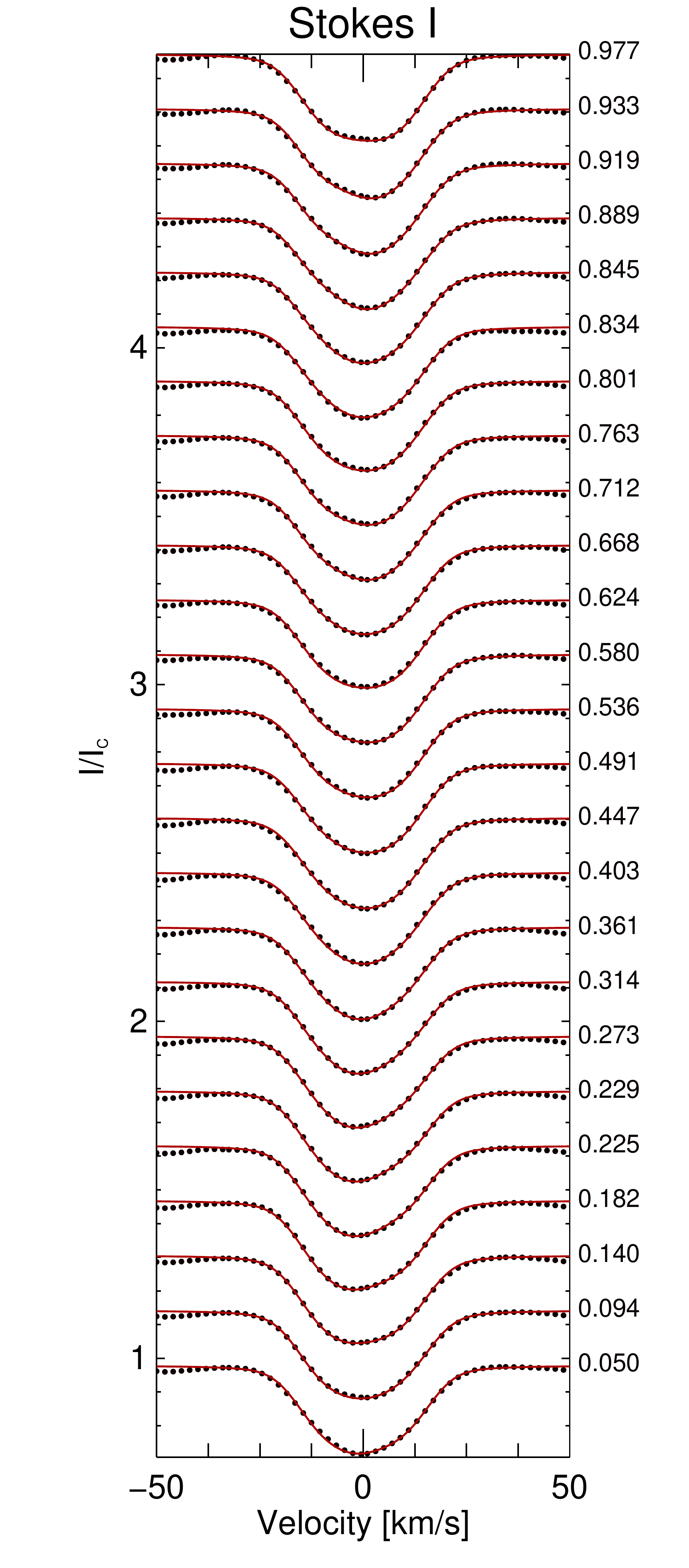}&
\hspace{-0.5cm} \includegraphics[scale=0.185,clip=true]{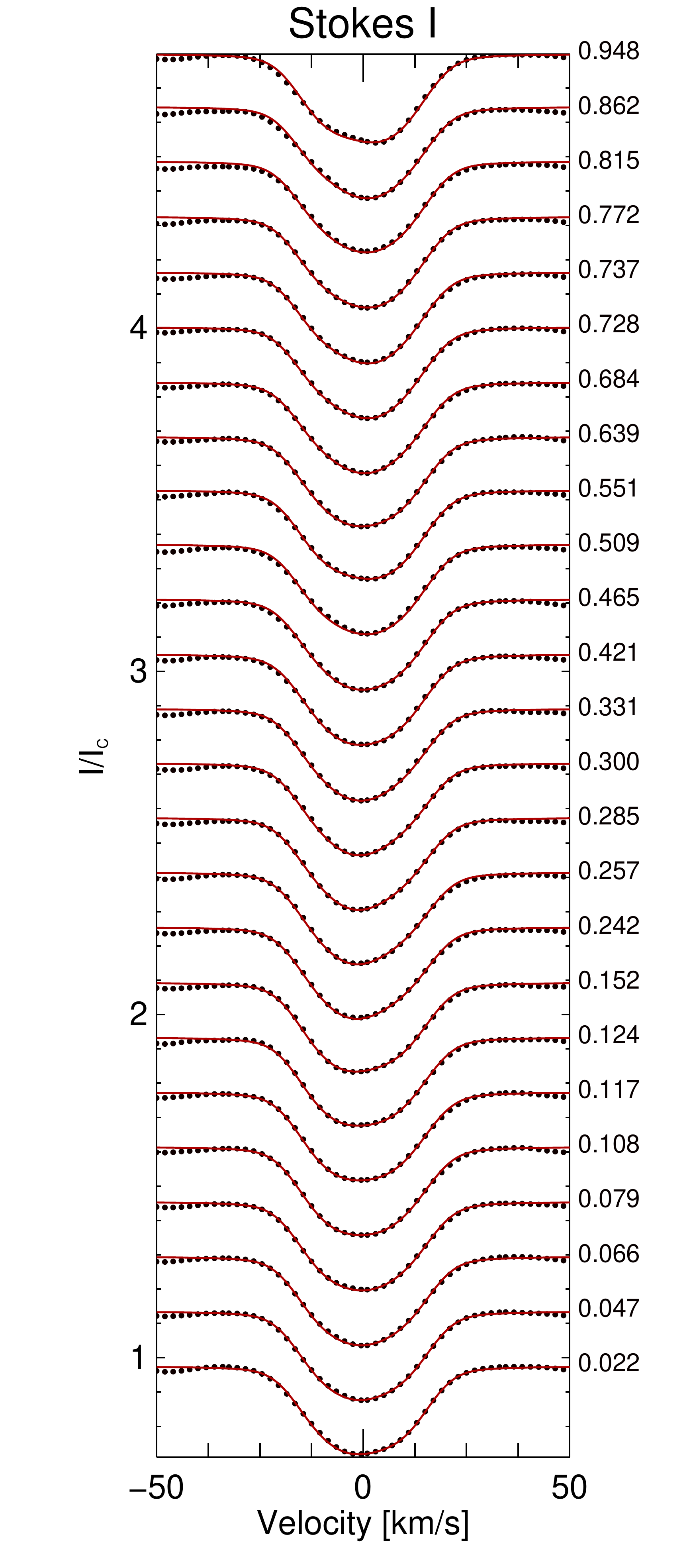}&
\hspace{-0.5cm} \includegraphics[scale=0.185,clip=true]{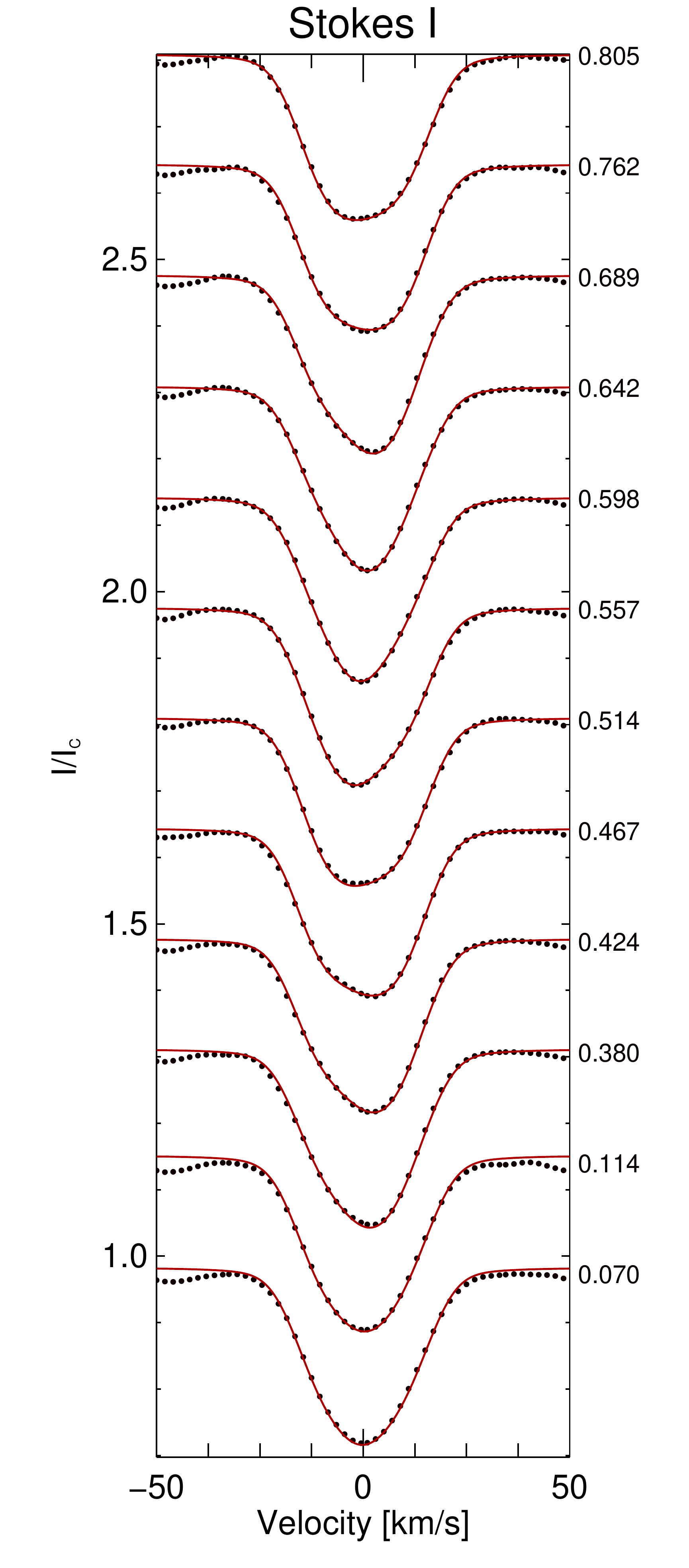}\\
{\small 2010 - set 6} & {\small 2010 - set 7} & {\small 2010 - set 8} & {\small 2010 - set 9} & 
{\small 2010 - set 10} \\
\includegraphics[scale=0.185,clip=true]{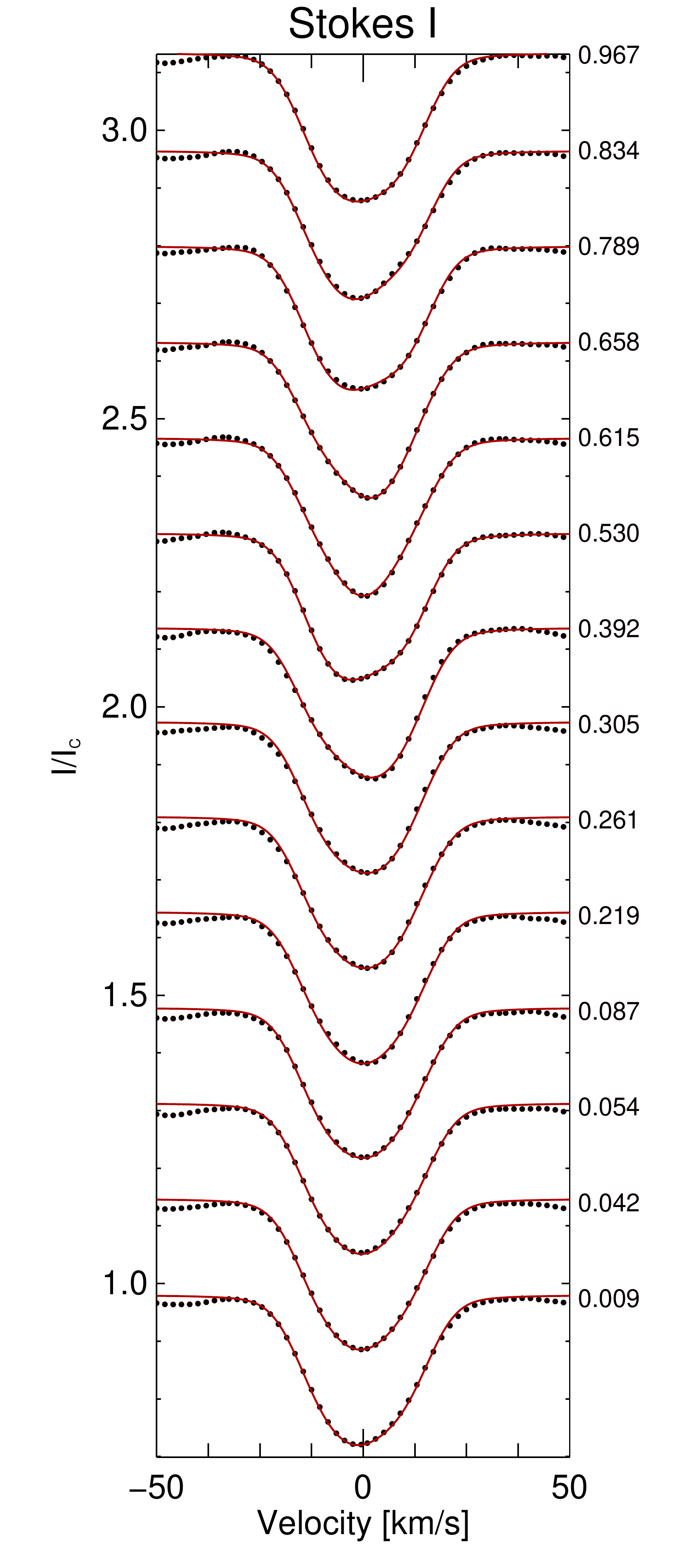}&
\hspace{-0.5cm} \includegraphics[scale=0.185,clip=true]{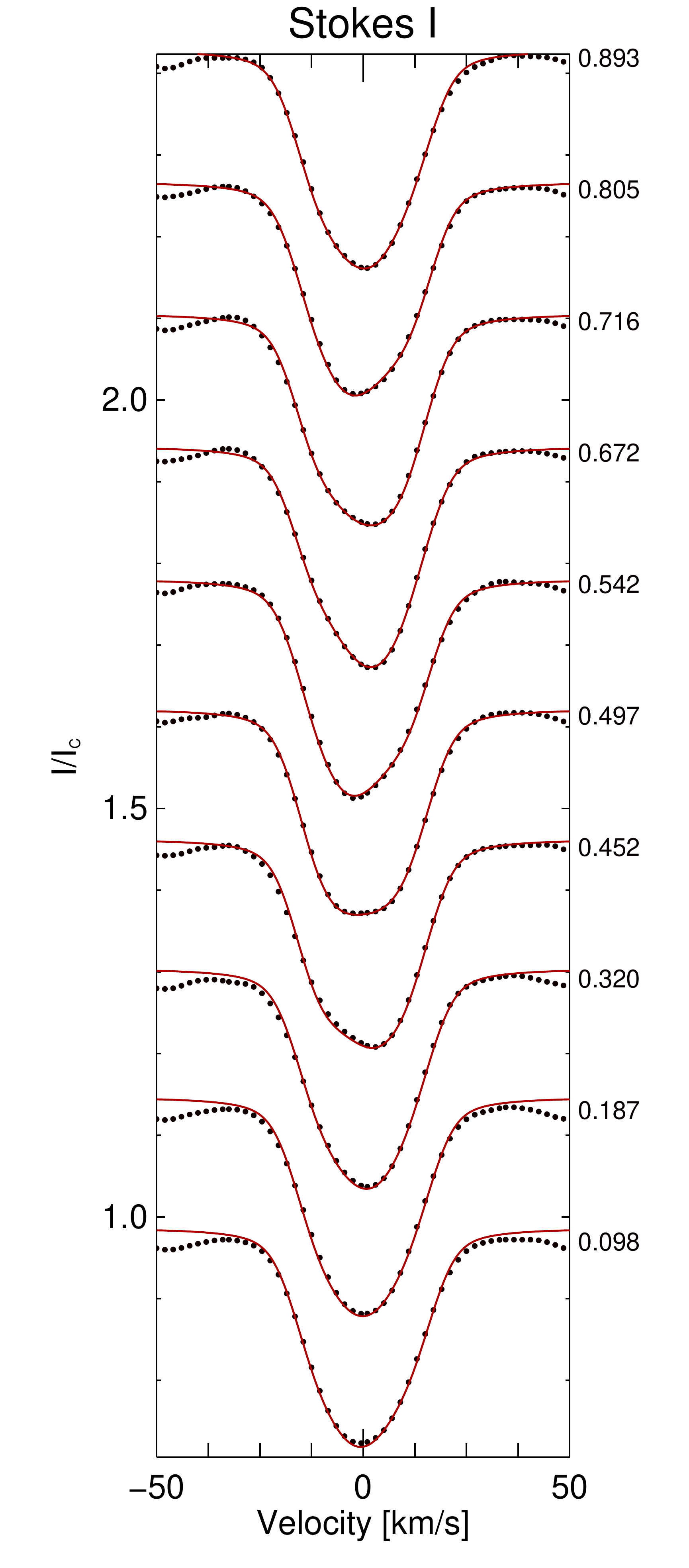}&
\hspace{-0.5cm} \includegraphics[scale=0.185,clip=true]{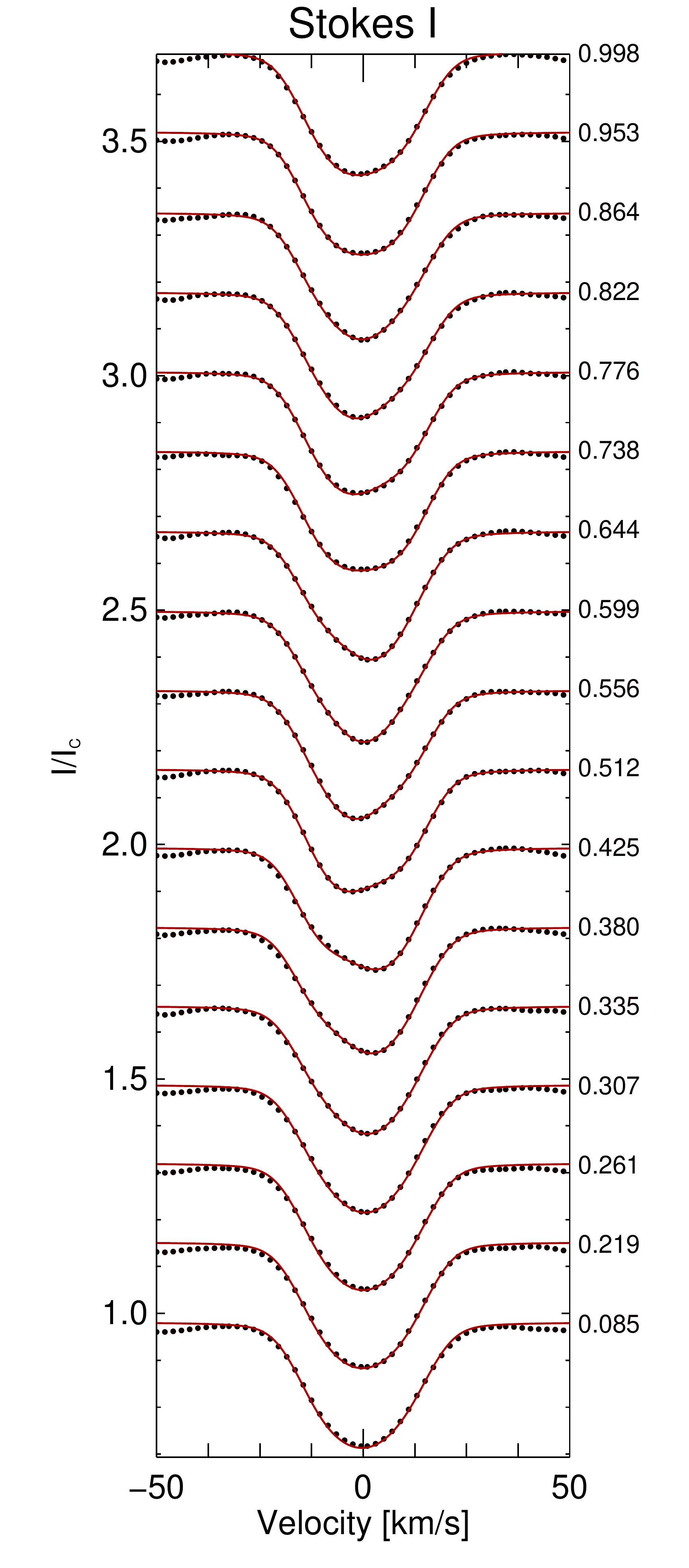}&
\hspace{-0.5cm} \includegraphics[scale=0.185,clip=true]{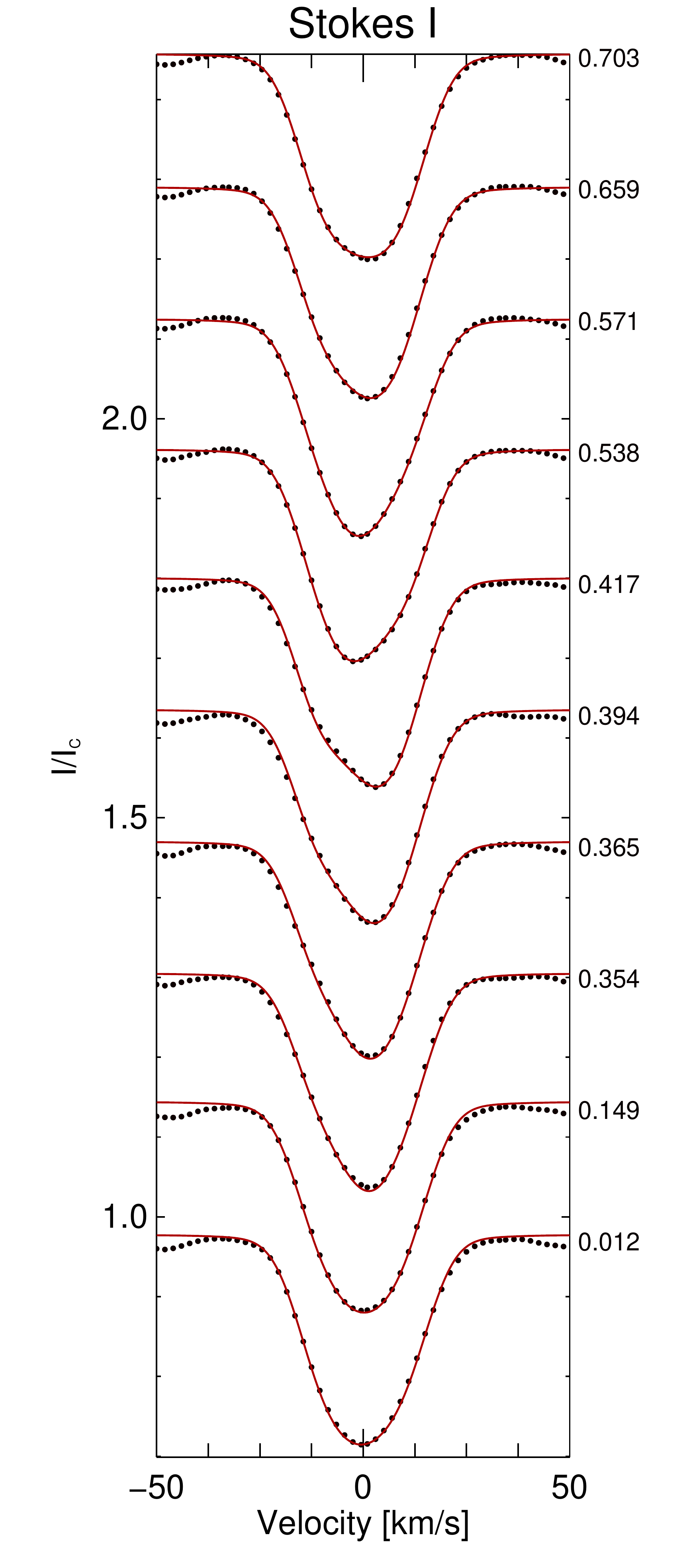}&
\hspace{-0.5cm} \includegraphics[scale=0.185,clip=true]{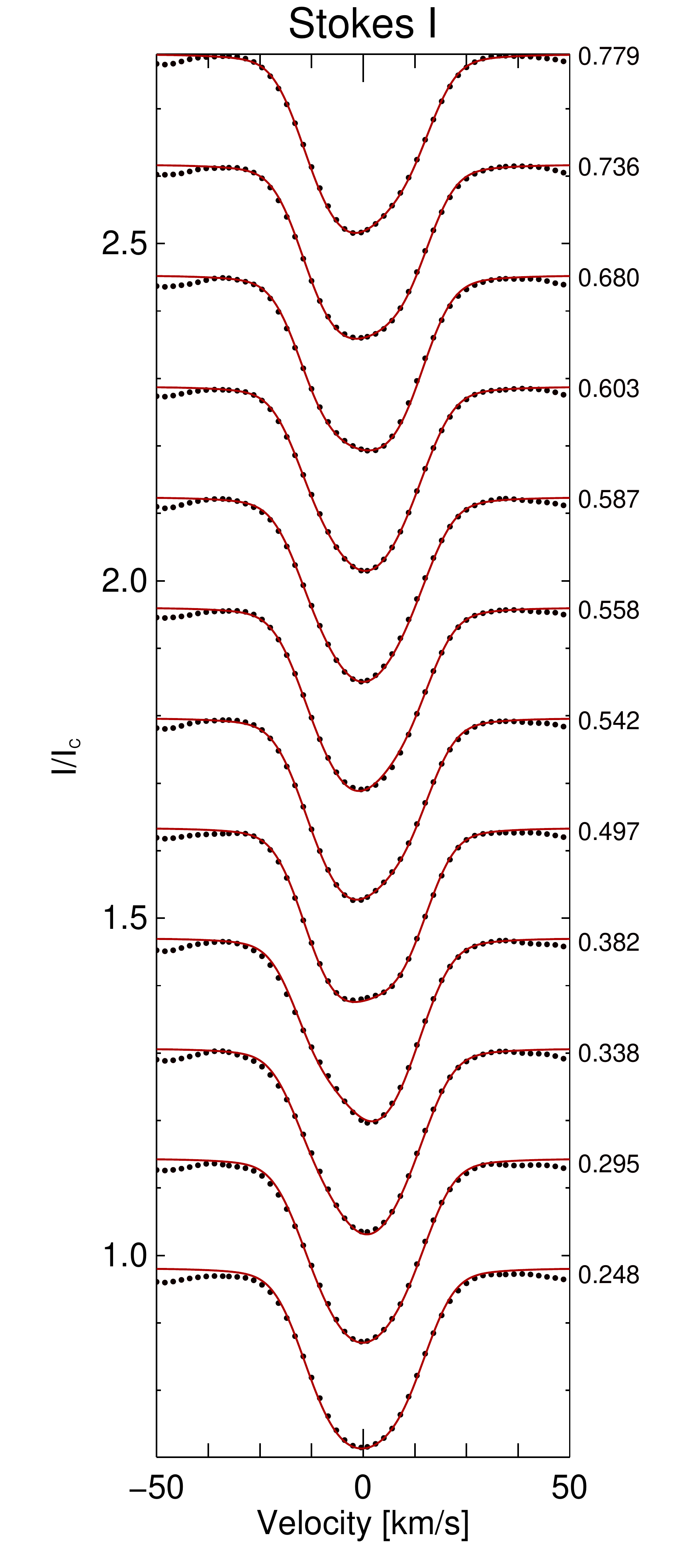}\\
{\small 2011 - set 11} & {\small 2011 - set 12} & {\small 2011 - set 13} & & & \\
\includegraphics[scale=0.185,clip=true]{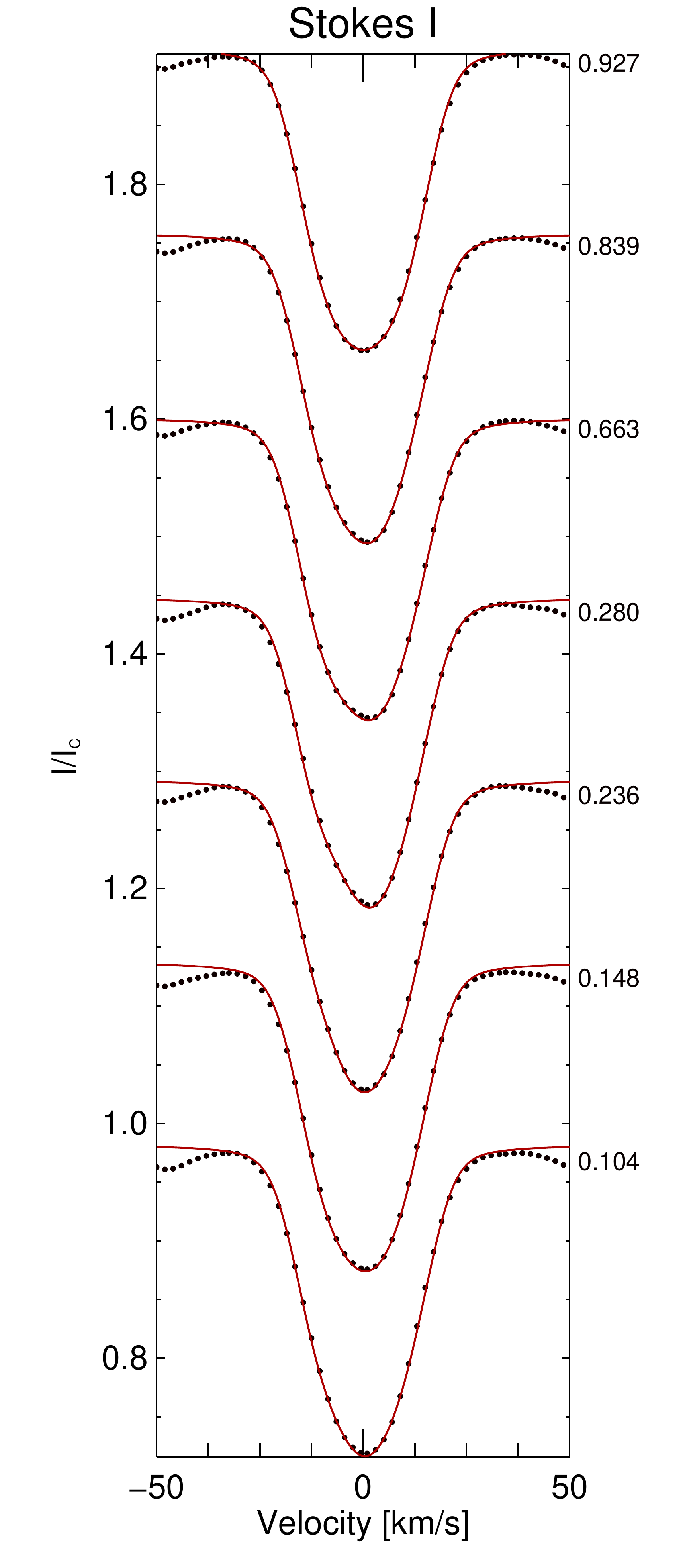}&
\hspace{-0.5cm} \includegraphics[scale=0.185,clip=true]{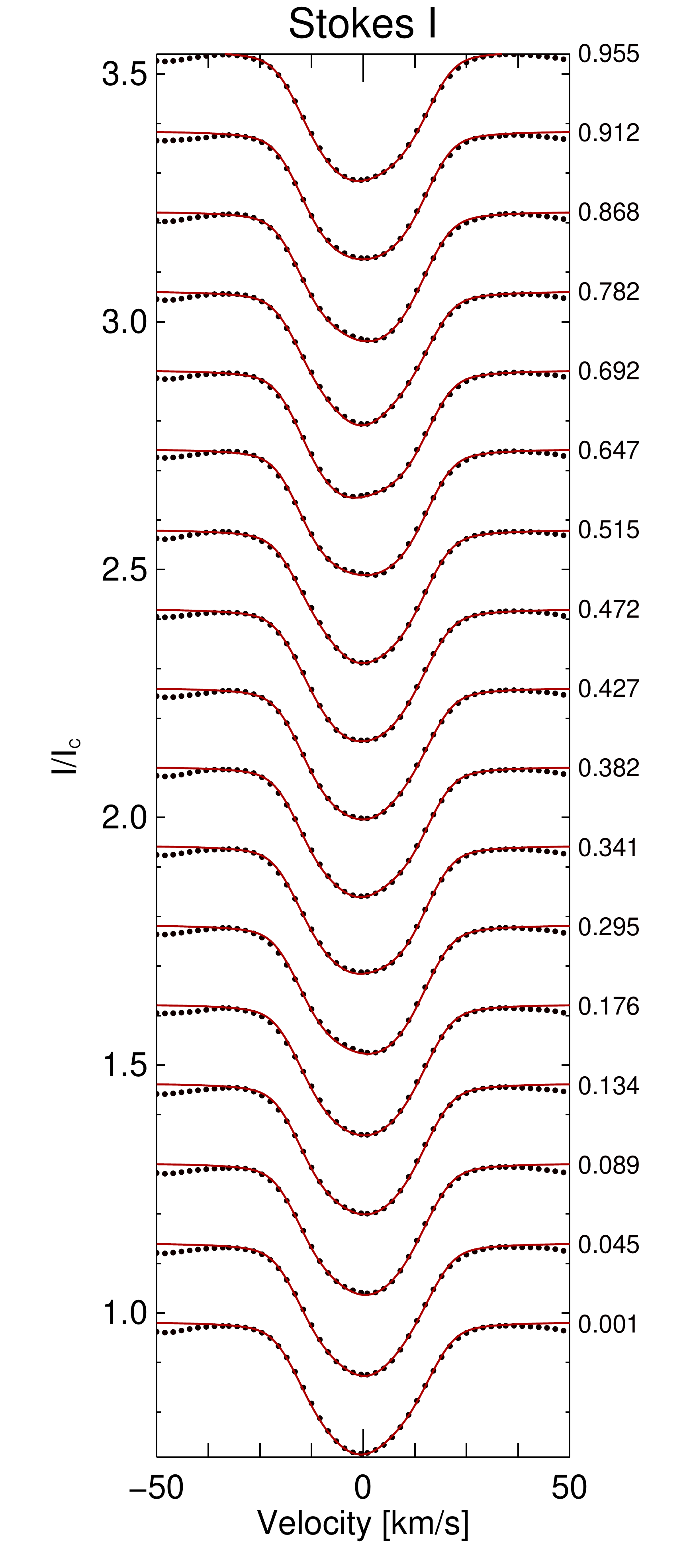}&
\hspace{-0.5cm} \includegraphics[scale=0.185,clip=true]{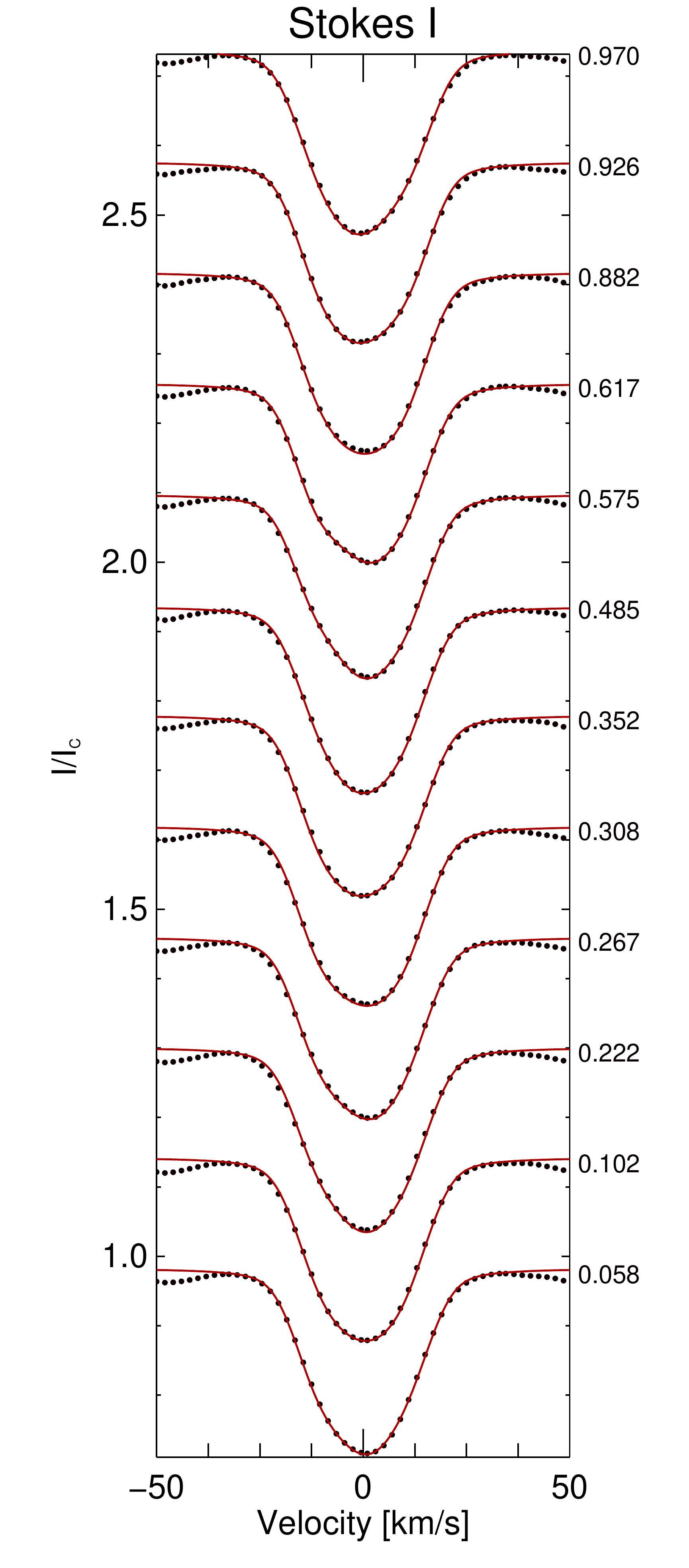}\\
\end{tabular}
\end{center}
\caption{Observed (dots) and inverted (continuous curve) line profiles
of 13 sets. For a given set, the corresponding orbital phase of each line 
profile is shown at the right side of the related panel. Observing season
and set numbers are shown at the top of the panels.}\label{A1}
\end{figure*}

Fig.~\ref{A1} shows observed and inverted line profiles of each 
Doppler image given in Figs.~\ref{F6} to \ref{F8}.

\end{appendix}

\end{document}